\documentclass[twocolumn,amssymb,fleqn]{revtex4-2} 
\usepackage{epsfig,amssymb,amsmath,graphicx,hyperref,subfigure}  
\usepackage{color}

\usepackage{multirow} 
\usepackage{tabularx}

\usepackage{graphicx}
\graphicspath{{img/}}

\newcommand{\be} {\begin{eqnarray}}
\newcommand{\ee} {\end{eqnarray} }

\newcommand{\f} {\frac }
\newcommand{\la} {\langle }
\newcommand{\ra} {\rangle }

\newcommand{\lb} {\left( }
\newcommand{\rb} {\right) }

\begin{document}

\title{A molecular dynamics simulation of thermalization of crystalline lattice
with harmonic interaction} 
\author{Zhenwei Yao}
\email{zyao@sjtu.edu.cn}
\affiliation{School of Physics and Astronomy, and Institute of Natural
Sciences, Shanghai Jiao Tong University, Shanghai 200240, China}
\begin{abstract} 
  Understanding the realization of thermal equilibrium through the thermalization
  process in a many-body system is a fundamental and complex scientific
  question, bridging thermodynamics and classical dynamics and connecting to a
  host of physical phenomena, such as mechanical instabilities in a thermal
  environment. In this work, based on the harmonic lattice model, we investigate
  the thermalization process in both velocity and coordinate spaces, by
  examining microscopic dynamics on the atomic level. We show the distinct
  relaxation rates of the transverse and longitudinal components of the
  velocity, reveal the power law governing the nonlinear proliferation of
  dominant frequencies, and observe the concurrent rapid proliferations of
  frequencies and topological defects. We also show that the lattice system's
  persistent out-of-plane deformations exhibit two-stage fluctuation behaviors,
  characterized by distinct power laws of fractional exponents and associated with
  the broken up-down symmetry. This work demonstrates the rich dynamics
  underlying the thermalization process, and advances our understanding on the
  dynamical adaptations of many-body systems to external disturbances.  
\end{abstract}

\maketitle

\section{Introduction}

Thermal equilibrium is a remarkable state admitted by many-body systems
containing many degrees of
freedom~\cite{gibbs2014elementary,aleksandr1949mathematical,Landau1999a,krylov2014works}.
Persistent thermal fluctuations lead to the formation of exceedingly rich
spatiotemporal
structures~\cite{prigogine1971,Nelson2004c,galliano2023two} and even
cause singular collective behaviors manifested in phase
transitions~\cite{Kosterlitz1973,elementsNishi}.  Inquiry into the fundamental
question on the realization of the thermal equilibrium state via the
thermalization process can be traced back to the pioneers in the development
of the subject of statistical
mechanics~\cite{maxwell1860v,boltzmann1964lectures,ehrenfest2002conceptual,gibbs2014elementary}.
A key question is how the system is thermalized, losing the information of 
initial state and exploring permissible states in
time~\cite{sinai1989dynamical,prigogine2017non,dumas2014kam,krylov2014works}.
Explorations into the thermalization problem deepen our understanding on the
fundamental concepts of ergodicity, reversibility, and entropy, which are
essential for understanding macroscopic
processes~\cite{ehrenfest2002conceptual,callen1951irreversibility,zheng1996ergodicity,frenkel2015order,wang2024thermalization}.
On the macroscopic level, the relaxation toward thermal equilibrium has been
examined in the thermodynamic framework of force and
flux~\cite{onsager1931reciprocal,prigogine1971,kubo1966}.

The perspective of nonlinear dynamics yields new insights into the microscopic
mechanism of the thermalization
process~\cite{li1975period,scheck2010mechanics,kaplan2012understanding,trachenko2015collective}.
Specifically, the discovery of the deterministic chaos through an infinite
sequence of period-doubling bifurcations sheds light on the origin of the
randomness arising in the thermalization
process~\cite{scheck2010mechanics,kaplan2012understanding,PhysRevLett.122.024102,PhysRevE.101.032211,yao2022collective,yao2023non}.
As such, the thermalization phenomenon presents a rich and complex problem that
bridges thermodynamics and classical dynamics, and it has inspired profound
discussions on the physics on both macroscopic and microscopic
levels~\cite{kadanoff1986two,Kadanoff1999,trachenko2015collective}. The variety
of interaction potentials and confining geometries enriches the problem and
imposes a challenge to fully understanding the thermalization
process~\cite{fermi1955studies,dauxois2002dynamics,berman2005fermi,gallavotti2007fermi,mulansky2009dynamical,ribeiro2014ergodicity,horn2014does,NelsonP2016,zaccone2023theory,jonay2024physical,vargas2025order}.
The harmonic crystalline lattice system offers a suitable platform for studying
the thermalization problem in a many-body system, due to the simplicity of its
interaction potential and the geometric nonlinearity arising in the lattice
configuration that breaks the integrability of the system~\cite{de2009relating,PhysRevLett.122.024102,PhysRevE.101.032211}.
The harmonic lattice system also provides a general model for diverse
2D regular particle packings near mechanical equilibrium, where the physical
interaction can be approximated by a harmonic potential~\cite{keim2004harmonic}.
Furthermore, the inquiry into the microscopic fluctuation dynamics within
the harmonic lattice system represents an extension of classical elasticity
into the thermodynamic regime~\cite{Landau1986,audoly2010elasticity}.
Elucidating the thermalization process in the crystal system has a strong
connection to a host of important physical phenomena, such as the mechanical
instabilities of crystalline materials in a thermal environment and 2D crystal
melting as a prototypical topological phase
transition~\cite{halperin1978theory,strandburg1988two,nelson2002defects}.

The goal of this work is to explore the thermalization process from the dynamic
viewpoint within the context of the harmonic lattice model, focusing on the
nonlinear dynamics and statistical regularity. The model consists of a harmonic
triangular lattice of circular shape with anchored boundary, constituting a
drum-like configuration.  Note that under the similar clamped
boundary condition, the rich interplay of thermal fluctuations and elasticity
has been investigated in crystalline ribbon and sheet
systems~\cite{Nelson2004c,blees2015graphene,wan2017thermal,le2021thermal,hanakata2021thermal,chen2022spontaneous}.
In our drum model, the dynamics is introduced by imposing a random velocity
vector on each particle in the initial state. The subsequent evolution of the
lattice conforms to the Hamiltonian dynamics. The numerical approach based on
the Verlet integration algorithm allows us to track the trajectory of each
particle on the atomic level~\cite{rapaport2004art}. The crystalline lattice as
a dissipationless Hamiltonian system is subject to persistent fluctuations upon
the initial disturbance. The dynamics of the thermalization process is analyzed
in both velocity and coordinate spaces. We also perform spectral analysis of the
kinetic energy curve, aiming at seeking the statistical regularity underlying
the complex dynamic evolution of the particle configuration.

The key results of this work are presented below. (1) We first show the
significantly faster relaxation of the longitudinal component of the velocity
compared with that of the transverse component, both of which ultimately
converge to the Maxwell distribution of common temperature that is proportional
to the squared strength of initial disturbance. (2) A power law governing the
nonlinear proliferation of dominant frequencies is revealed; the exponent is
tunable by the disturbance strength. A discrete theoretical model is proposed to
explain the nonlinear feature in the frequency proliferation process.  (3)
Analysis of the drum system's persistent in-plane and out-of-plane fluctuations
reveals the concurrent rapid proliferations of frequencies and topological
defects, and the two-stage fluctuation behaviors associated with the broken
up-down symmetry, respectively.

\section{Model and Method}

\begin{figure}[t]  
\centering 
\includegraphics[width=3.1in]{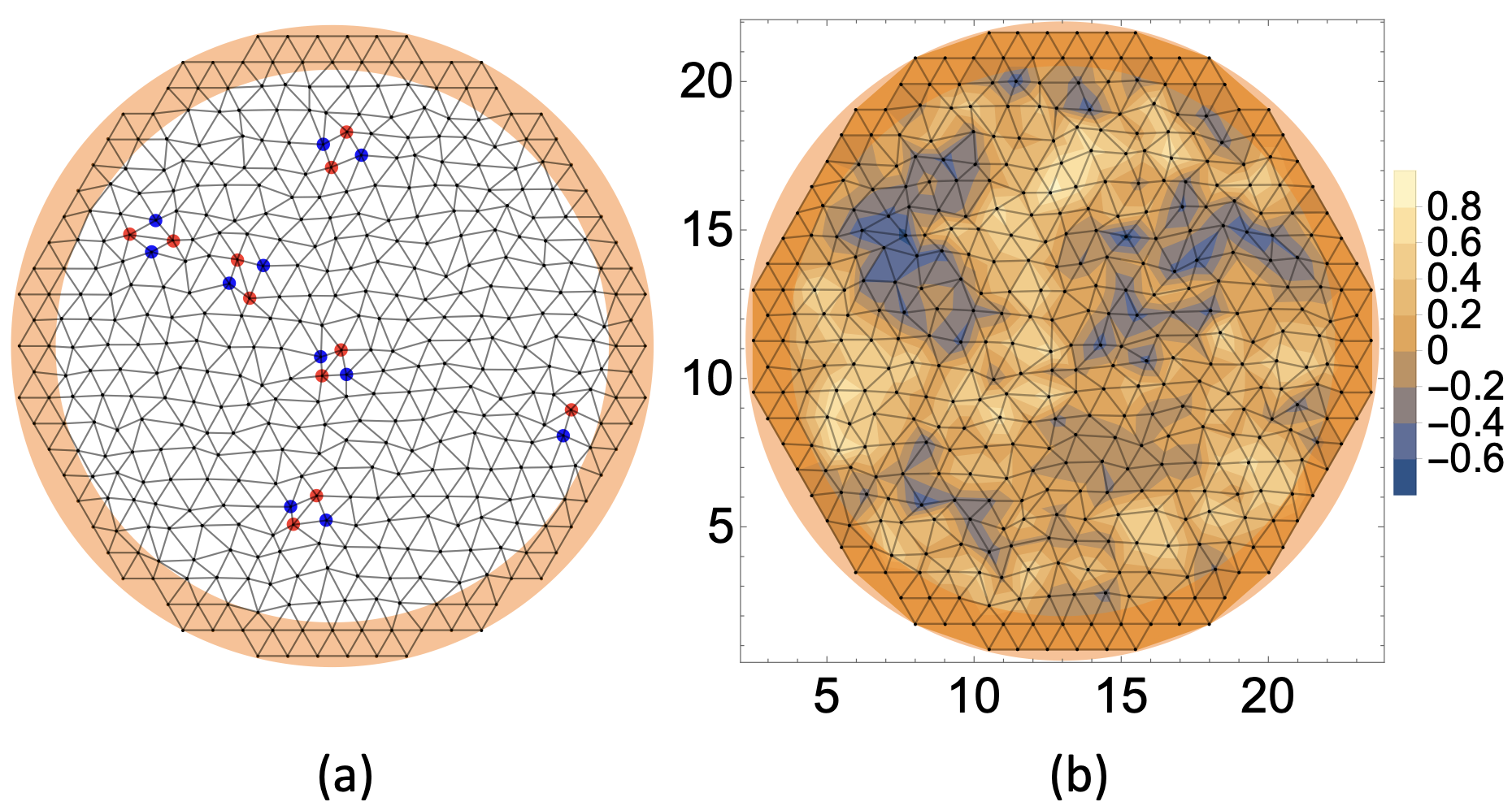}
  \caption{Illustration of the drum model with the anchored boundary (in
  orange). (a) The drum model consists of the triangular lattice of linear
  springs (bonds) and point particles (vertices). The five- and seven-fold
  disclinations are colored in red and blue, respectively. Disclinations are
  identified by the Delaunay triangulation of particle configuration projected
  onto the plane. Note that the geometric bonds established by the Delaunay
  triangulation algorithm (not shown in the figure) can be distinct from the
  physical bonds (springs) connecting particles. (b) The contour plot shows the
  instantaneous out-of-plane deformation in the dynamical evolution of the
  lattice.  $t=5\tau_0$.  $v_0=0.5$. $N=316$.
  }
\label{model}
\end{figure}

The model consists of a harmonic triangular lattice of circular shape with
anchored boundary, constituting a drum-like configuration; see Fig.~\ref{model}.
In the drum model, the $N$ particles are connected by identical linear springs
of stiffness $k_0$ and rest length $\ell_0$, and they are allowed to move in 3D
space. Note that we also consider the case where the motion of particles is
confined on the plane of the drum (denoted as the $x-y$ plane). To implement the
anchored boundary condition, the particles within the annulus (in orange) are
fixed in space.  The lattice is initially in mechanical equilibrium; the lattice
spacing is equal to $\ell_0$, the rest length of the spring. In this work, the
length, mass, and time are measured in units of $\ell_0$, $m_0$ (the mass of the
particle), and $\tau_0$; $\tau_0=\sqrt{m_0/k_0}$.

The dynamics is introduced by imposing an initial disturbance to the drum.
Specifically, we specify a randomly distributed velocity to each particle:
$\vec{v}_{\rm{ini}} = v_0 \vec{\xi}$, where $\vec{\xi}$ is a random vector in 3D
space. The three components of $\vec{\xi}$ are independent, and they conform to
the uniform distribution in the range of $[-1, 1]$. The strength of the
disturbance is indicated by $v_0$, which is a key tuning parameter in our model.
Upon the initial disturbance, the motion of the particles conforms to
Hamiltonian dynamics. The Hamiltonian of the system is:
\be
H = \sum_{i\in V} \frac{\vec{p}_i^2}{2m_0} + \sum_{\alpha \in E}
V_0(r_{\alpha}),
\ee
where the summation is over all the particles ($V$) and bonds ($E$).
$\vec{p}_i$ is the momentum of particle $i$. $r_{\alpha}$ is the length of
spring $\alpha$. The harmonic potential created by the linear spring of
stiffness $k_0$ and rest length $\ell_0$ is
\be
V_0(r_{\alpha}) =\f{1}{2} k_0(r_{\alpha}-\ell_0)^2. \label{harmonic}
\ee
This work focuses on the harmonic case. The dynamical effects of nonlinear
interactions are also examined (see Appendix A for details).

We solve for the coupled equations of motion by the standard Verlet integration
method~\cite{rapaport2004art}. The time step $dt=10^{-3}$, under which the total
energy is well conserved. Specifically, for the typical lattice consisting of
847 particles, the relative variation of the total energy is at the order
$10^{-4}$ during five million simulation steps. The obtained temporally-varying
particle configurations are analyzed from the multiple perspectives of velocity
distribution, spectral structure, and particle position fluctuations.

\section{Results and discussion}

This section consists of four subsections. In Sec. III A, we discuss the
thermalization process in the velocity space. We show the distinct relaxation
rate of the longitudinal and transverse components of the velocity, both of
which ultimately converge to the Maxwell distribution of common temperature.  In
Sec. III B, we perform spectral analysis of the thermalization process and
reveal the power law in the proliferation of dominant frequencies with the
extension of the observation time; the exponent is tunable by the disturbance
strength. A discrete theoretical model is proposed in Sec. III C for explaining
the nonlinear feature in the frequency proliferation process. In Sec.  III D, we
examine the thermalization process in the coordinate space. Analysis of the
in-plane and out-of-plane deformations of the lattice reveals two key findings:
the concurrent rapid proliferations of frequencies and topological defects, and
the two-stage fluctuation behaviors with the increase of the disturbance
strength, respectively.

\begin{figure}[t]  
\centering 
\includegraphics[width=3.5in]{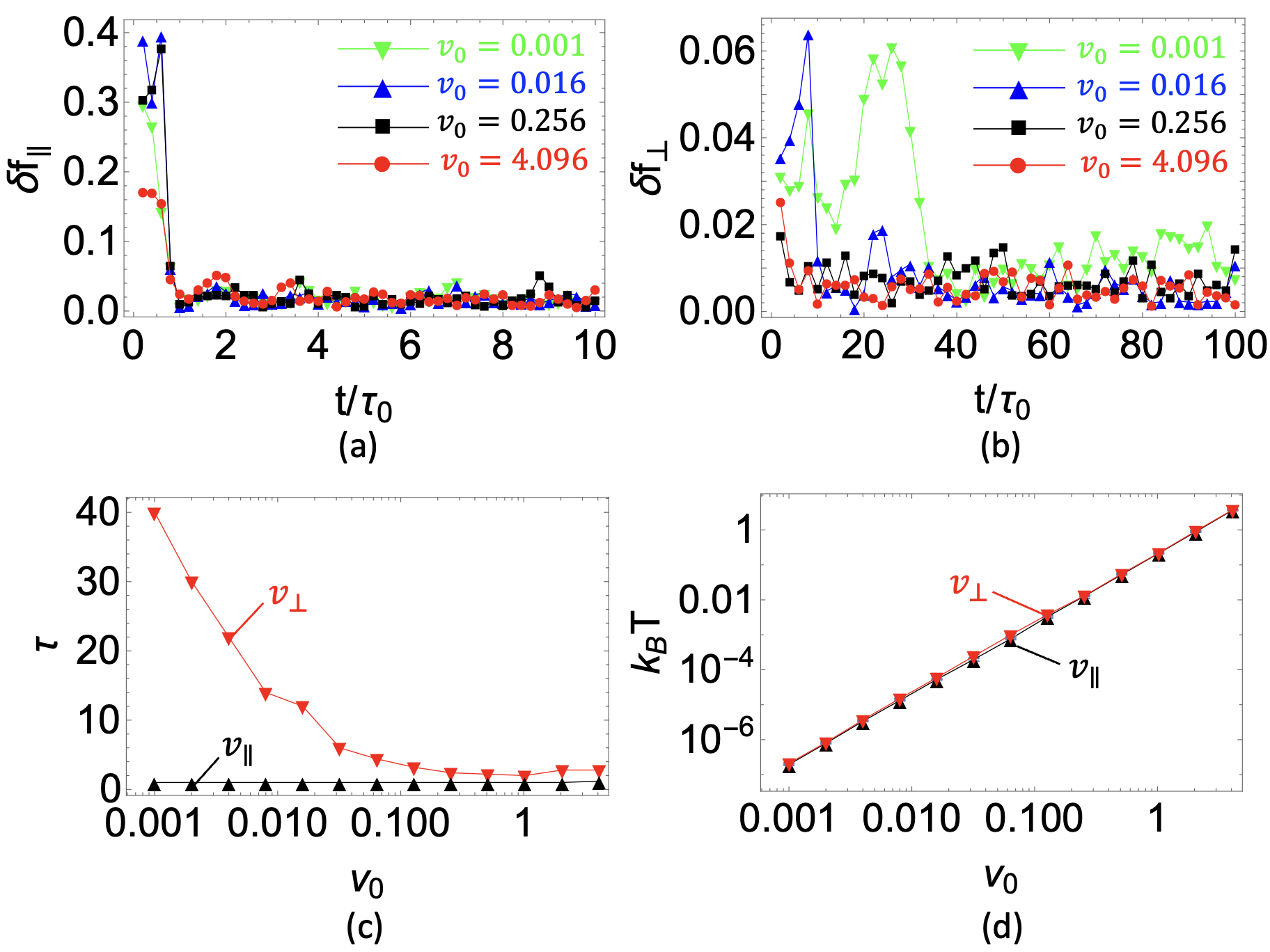}
  \caption{Characterization of the thermalization process by the relaxation of
  velocities. (a) and (b) Plots of $\delta f_{\parallel}$ and $\delta f_{\perp}$
  (the deviations of $v_{\parallel}$- and $v_{\perp}$-distributions from the
  Maxwell distributions) versus time at typical values of $v_0$. (c) Distinct
  relaxation time of $v_{\parallel}$ and $v_{\perp}$ versus $v_0$. (d) The
  log-log plot of temperature (as derived from the Maxwell distributions) versus
  $v_0$ show that $T\propto v_0^2$ for both cases of $v_{\parallel}$ and
  $v_{\perp}$. The sampling is during $t\in [900\tau_0, 1000\tau_0]$ at the time
  resolution $\Delta t=10\tau_0$. The standard deviations are indicated by the
  error bars. $N=1406$. 
  }
\label{velocity}
\end{figure}

\subsection{Relaxation of velocities}

We first analyze the evolution of the velocity distribution upon the initial
disturbance of varying strength. Specifically, we separately analyze the
distributions of both longitudinal ($v_{\parallel}$) and transverse
($v_{\perp}$) components of the velocity. The $v_{\parallel}$ and $v_{\perp}$
components correspond to the in-plane and out-of-plane motions of the
particles.

The relaxation process is characterized by the temporal variation of $\delta
f_{\alpha}(t)$, the deviation of the instantaneous $v_{\alpha}$-distribution
from the corresponding Maxwell distribution. $v_{\alpha}=v_{\parallel}$ or $v_{\perp}$. The
quantity $\delta f_{\alpha}$ is defined as follows:
\begin{eqnarray}
  \delta f_{\alpha}(t) = \frac{\int [f_t(v_{\alpha})-f_0(v_{\alpha})]^2
  dv_{\alpha}}{\int f_0(v_{\alpha})^2 dv_{\alpha}},
\end{eqnarray}
where $f_t(v_{\alpha})$ is the instantaneous distribution of $v_{\alpha}$ at
time $t$, and $f_0(v_{\alpha})$ is the Maxwell distribution.
Figures~\ref{velocity}(a) and \ref{velocity}(b) show the plots of $\delta
f_{\parallel}$ and $\delta f_{\perp}$ versus time at typical values of $v_0$,
the strength of the initial disturbance. We observe the convergence of both
$v_{\parallel}$ and $v_{\perp}$ toward the Maxwell distribution, but at
significantly different rates.

The dependence of the relaxation time on $v_0$ is summarized in
Fig.~\ref{velocity}(c). $\tau_{\parallel}$ and $\tau_{\perp}$ refer to
the relaxation time of the $\delta f_{\parallel}(t)$ and $\delta f_{\perp}(t)$
curves, respectively. In simulations, both $\delta f$ curves decline and
eventually enter a zone of steady fluctuation. The relaxation time is defined as
the time when the $\delta f$ curves reach the first minimum within the
fluctuation zone, where $\delta f$ is within $2\%$ in general. In contrast to
the short, constant relaxation time $\tau_{\parallel}$ of the
$v_{\parallel}$-distribution, the relaxation time $\tau_{\perp}$ of the
$v_{\perp}$-distribution is significantly longer, especially in the small $v_0$
regime. The much faster relaxation of the $v_{\parallel}$-distribution indicates
that in-plane particle motion leads to a much higher efficiency of velocity
relaxation than out-of-plane motion.  Figure~\ref{velocity}(c) also shows that
increasing the strength of disturbance significantly reduces the value of
$\tau_{\perp}$ and accelerates the relaxation of $v_{\perp}$.  The
discrepancy in the fluctuations of orthogonal components of relevant quantities
(force and displacement) is also reported in both 2D and 3D disordered crystals
with athermal fluctuations caused by particle-size polydispersity; fluctuations
in forces orthogonal to the lattice directions are highly constrained as
compared to the fluctuations along lattice
directions~\cite{acharya2020athermal,maharana2022athermal}.

Despite the discrepancy in the relaxation time, both $v_{\parallel}$- and
$v_{\perp}$-distributions ultimately converge to the Maxwell distribution of
common temperature. Here, the concept of temperature, as defined in terms of the
dispersion of the Maxwell distribution, arises in the mechanical lattice system.
It is noticed that even under very small perturbation ($v_0=0.001$), the system
evolves toward thermal equilibrium as characterized by the full relaxation of
the velocity distribution; the corresponding temperature is at the order of
$10^{-7}$.  In the log-log plot in Fig.~\ref{velocity}(d), the linear relation
demonstrates a power law between $T$ and $v_0$.  Specifically, the
data for both cases of $v_{\parallel}$ and $v_{\perp}$ can be well fitted by the following
function:
\be
\log(k_BT) = b_{\alpha} + k_{\alpha} \log(v_0),
\ee
where the slopes of the $v_{\parallel}$- and $v_{\perp}$-lines in the log-log plot
in Fig.~\ref{velocity}(d) are $k_{\parallel}=2.007$ and $k_{\perp}=1.998$,
respectively.  $b_{\parallel}=-0.685$, and $b_{\perp}=-0.653$. To conclude, we
obtain the relation between the temperature $T$ and the strength of the initial
disturbance $v_0$ within the tolerance of error:
\be
T\propto v_0^2. \label{T_v0}
\ee

\begin{figure*}[t]  
\centering 
\includegraphics[width=6in]{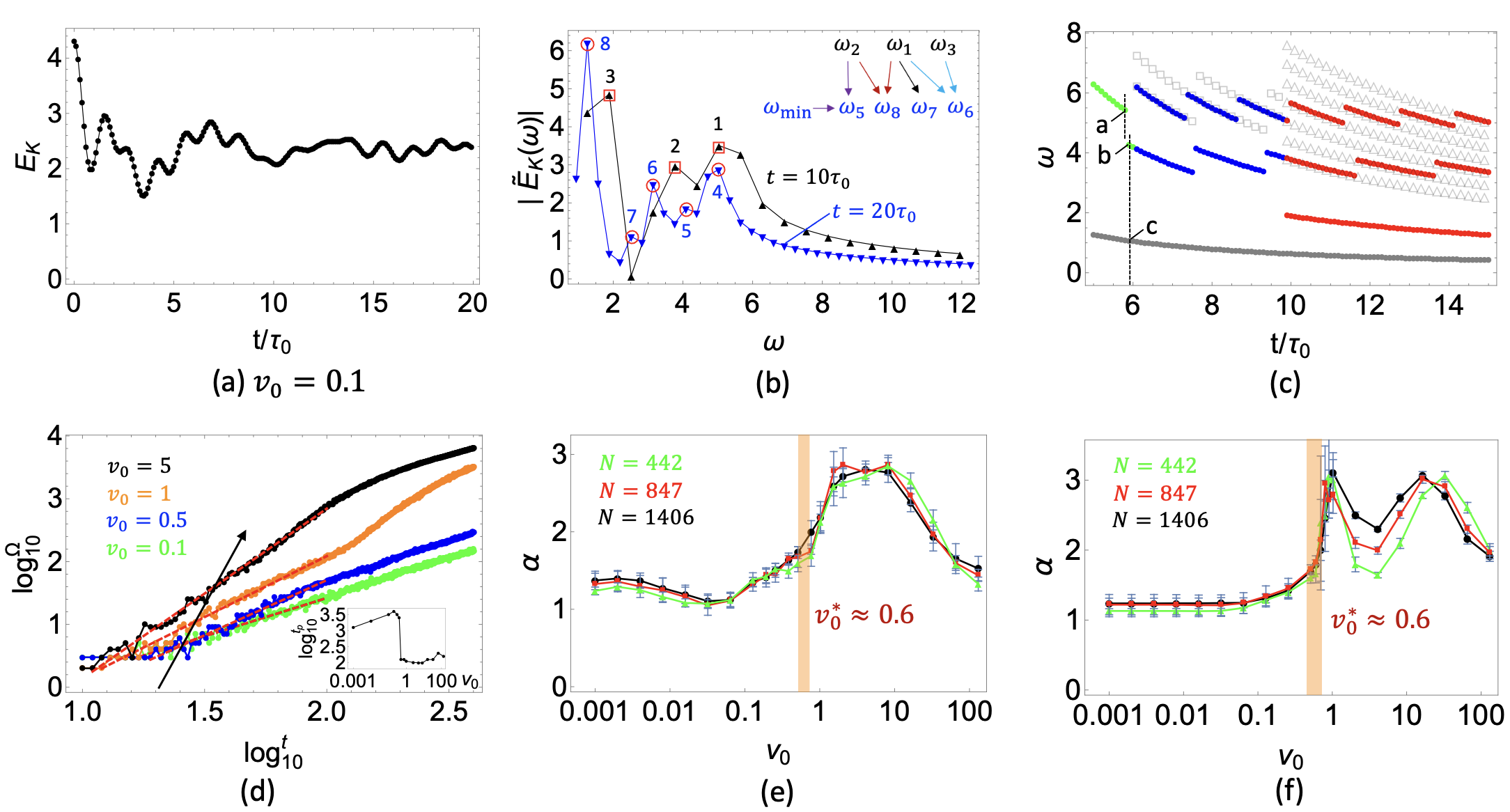}
  \caption{Thermalization process from the perspective of the nonlinear proliferation of
  dominant frequencies. (a) Plot of the kinetic energy versus time. (b) The
  frequency spectra of kinetic energy for the observation time of $t=10\tau_0$
  and $t=20\tau_0$.  (c) The number of dominant frequencies increases from one
  (green), two (blue) to three (red) with the extension of the observation time
  $t$. The empty square and triangle symbols show the sum of the lower frequency
  in both two- and three-frequency regimes and a multiple of $\omega_{\rm{min}}$
  (the lowest curve in gray).  (d) The growth of the total number of frequencies
  [$\Omega(t)$] in the early stage conforms to the power law. (e) and (f)
  Dependence of the exponent $\alpha$ in the power law of $\Omega(t)$ on $v_0$
  for the cases of 2D lattices allowing and restricting out-of-plane
  deformations, respectively. The arrows indicate the critical value of $v_0$,
  above which the $\alpha(v_0)$ curve experiences a rapid rise.  The statistical
  analysis is based on 10 independent initial distributions of velocity. The
  sampling is during $t\in [\tau_0, 100\tau_0]$ at the time resolution $\Delta
  t$. $\Delta t=0.001\tau_0$ in (c); otherwise, $\Delta t=0.01\tau_0$.
  $v_0=0.1$ in (a)-(c).  $N=847$ in (a)-(d). 
  }
\label{frequency}
\end{figure*}

\subsection{Power law in the proliferation of frequencies}

In the preceding subsection, we discuss the thermalization of the disturbed drum
system in the velocity space by examining the velocity relaxation. Here, we
further analyze the thermalization process from the perspective of the dynamical
modes. Spectral analysis offers a powerful tool for extracting featured
dynamical and even structural patterns, as demonstrated in recent studies of
amorphous solids and nonlinear dynamical
systems~\cite{xu2007excess,manning2011vibrational,PhysRevLett.122.024102,yao2023non,wu2023topology,zaccone2023theory}.
In the drum model, the geometric nonlinearity of the triangular lattice causes
the mixing of frequencies even after the distribution of velocity reaches
equilibrium. In this subsection, we aim at identifying the statistical
regularities that govern the complex dynamic evolution of the lattice, by
analyzing the proliferation of dominant frequencies with the extension of the
observation time.

In Fig.~\ref{frequency}(a), we show the variation of the kinetic energy with the
increase of the observation time $t$. To extract the frequency information, we
perform Fourier transformation of the kinetic energy curve $E_K(t)$. The Fourier
transformed kinetic energy curves in the observation times of $t=10\tau_0$ and
$t=20\tau_0$ are presented in Fig.~\ref{frequency}(b). The temporally-varying
dominant frequencies (at the marked peaks) offer a unique approach to
understanding the intricate dynamic evolution of the system. Note that the
frequency of the system is bounded.  $\omega_{\rm{min}}=2\pi/t$ and
$\omega_{\rm{max}}=2\pi/\delta t$, where $t$ is the observation time, and
$\delta t$ is the time resolution in sampling.  In the inset graph of
Fig.~\ref{frequency}(b), we show the relation of the dominant frequencies (which
are called frequencies later in this text).  Specifically, by the
frequency-mixing processes, $\omega_5=\omega_2+\omega_{\rm{min}}$, $\omega_8 =
\omega_1-\omega_2$, and $\omega_6 = \omega_1-\omega_3$. The frequency-halving
process leads to $\omega_7 = \omega_1/2$.

The variation of the dominant frequencies (in green, blue and red) and the
observation time $t$ is presented in Fig.~\ref{frequency}(c). The lowest curve
(in gray) shows the variation of $\omega_{\rm{min}}$;
$\omega_{\rm{min}}=2\pi/t$, which is determined by the total observation time
$t$.  With the increase of $t$ until $t=15\tau_0$, we see that the number of
dominant frequencies increases from one, two to three, as represented by the
green, blue and red curves, respectively. We also observe the gradual drift of
the frequency in time.  The observed frequency drift phenomenon
represents a nonlinear dynamic effect of the triangular lattice. The
nonlinearity of the triangular lattice originates from its geometric
configuration, which leads to anharmonic terms in the
potential~\cite{de2009relating,PhysRevLett.122.024102,PhysRevE.101.032211,yao2023non}.
Due to the intrinsic geometric nonlinearity, the triangular lattice exhibits
nonlinear dynamics even in the absence of external stimuli and under the
harmonic interaction.

Fig.~\ref{frequency}(c) shows that the proliferation of frequencies is realized
by the mixing of frequencies. For example, examination of the frequencies at
$t=5.78$ (at site $a$) and $t=5.88$ (at sites $b$ and $c$) shows that $\omega_a
= \omega_b + \omega_c$. Furthermore, it is found that the higher
frequencies in both two- and three-frequency regimes (indicated by blue and red
curves) are the combination of the lower frequency and a multiple of
$\omega_{\rm{min}}$, which are indicated by the empty square and triangle
symbols.

The proliferation of frequencies in even longer observation time is presented in
Fig.~\ref{frequency}(d). Fig.~\ref{frequency}(d) shows the variation of the
total number of frequencies [denoted as $\Omega(t)$] with the extension of the
observation time $t$ at typical values of $v_0$. The initially flat curves start
to rise for $t > t^*$.  Within the interval $t \in [t^*, t_p]$, the linear fits
(dashed red lines) in the log-log plots indicate that the $\Omega(t)$ curves are
well described by a power law: 
\be
\Omega(t) - \Omega(t^*) = C (t-t^*)^{\alpha}, \label{Omega1}
\ee
where the value of the exponent $\alpha$ is given by the slope of the linear
fit, and $C$ is some constant.  The dependence of $t_p$ on $v_0$ is
shown in the inset plot of Fig.~\ref{frequency}(d). We see an abrupt decline of
the $t_p$-curve during $v_0\in [0.4, 0.5]$; the value of $t_p$ is
reduced from $t_p=10^{3.4}\approx 2512$ at $v_0=0.4$ to $t_p=10^{2.1}\approx
126$ at $v_0=0.5$. The value of $t^*$ is relatively insensitive to $v_0$;
$t^*=16.8\pm 3.4$ for $v_0\in [0.001, 131]$.

According to Eq.(\ref{Omega1}), during $t \in [t^*, t_p]$, the growth of the
number of frequencies conforms to the following equation:
\be
\f{d}{dt}[\Omega(t) - \Omega(t^*)] = \alpha \f{\Omega(t) - \Omega(t^*)}{t-t^*},
\label{omega1}
\ee
which yields a power-law solution. The quantity $t-t^*$ in the denominator of the
right hand side term in Eq.(\ref{omega1}) describes the suppressed generation of
new frequencies in time. The upper bound of $\Omega$ is determined by the total
number of discretized time points within the observation time.

Here, we discuss the microscopic dynamical process of frequency proliferation
based on the power law of $\Omega(t)$. In time interval $\Delta t$, the total
number of frequencies is increased from $\Omega(t)$ to $n(t) \Omega(t)$. On
average, each frequency contributes $n(t)$ frequencies to $\Omega(t)$ in
$\Delta t$. According to Eq.(\ref{omega1}), 
\be
n(t) = 1 + \alpha \f{\Delta t}{t-t^*}\lb 1- \f{\Omega(t^*)}{\Omega(t)} \rb. 
\ee
Since $\Omega(t) \geq \Omega(t^*)$, $n(t)\geq 1$. For large $t$ (but still
within the $t \leq t_p$ power-law regime), $n(t) \approx 1 + \alpha \Delta t/t>1$.
The suppressed generation of new frequencies is characterized by the reduction
of $n(t)$ in time, scaling as $1/t$.

In Fig.~\ref{frequency}(e), we show the nonmonotonous dependence of the exponent
$\alpha$ on $v_0$ at typical system sizes. The value of $\alpha$ is insensitive
to the variation of system size. An important observation is the rapid rise of
the $\alpha(v_0)$ curve when $v_0$ exceeds some critical value $v_0^*$.
Specifically, the slope of the curve is significantly increased at
$v_0^*= 0.75$, $0.75$, and $0.5$ for the cases of $N=442$, $847$, and $1406$,
respectively. As such, the transition occurs at $v_0^*\approx 0.6$ as indicated
by the orange bar in Fig.~\ref{frequency}(e). The efficiency of frequency
proliferation, as measured by the value of $\alpha$, reaches maximum in the
range of $v_0 \in (2,8)$. As $v_0$ is varied over five orders of magnitudes
(from $v_0=0.001$ to $v_0=100$), the value of $\alpha$ ranges from 1.1 to 2.9.

To examine the robustness of the power law governing $\Omega(t)$ against
out-of-plane fluctuations, we also investigate the drum system with the motion
of particles restricted on the plane, and present the results in
Fig.~\ref{frequency}(f). Comparison of Figs.~\ref{frequency}(e) and
\ref{frequency}(f) shows that the mild undulation on the $\alpha(v_0)$ curve in
the regime of small $v_0$ is flattened with the suppression of the out-of-plane
fluctuations.  Similar to the case in Fig.~\ref{frequency}(e), we also
observe the rapid rise of the $\alpha(v_0)$ curve at $v_0^*\approx 0.6$ in
Fig.~\ref{frequency}(f). In the absence of out-of-plane fluctuations, the
highest efficiency of frequency proliferation occurs at the two peaks of the
$\alpha(v_0)$ curve in Fig.~\ref{frequency}(f), in contrast to the plateau in
Fig.~\ref{frequency}(e).  With the further increase of $v_0$, both $\alpha(v_0)$
curves in Figs.~\ref{frequency}(e) and \ref{frequency}(f) decline.  It indicates
the suppressed efficiency of frequency proliferation in the regime of strong
disturbance.

Here, it is of interest to extend the linear harmonic interaction to the
nonlinear regime. Simulations of the lattice system under the anharmonic
interaction show that the growth of the number of frequencies also conforms to
power law. More information is provided in Appendix A. The similarity of the
$\alpha(v_0)$ curves under the harmonic [Fig.~\ref{frequency}(e)] and anharmonic
[Fig.~\ref{nonlinear}(a)] interactions suggests that the nonlinear dynamics is
largely dictated by the intrinsic geometric nonlinearity of the triangular
lattice.

\subsection{Modeling nonlinear frequency proliferation processes}

\begin{figure}[t]  
\centering 
\includegraphics[width=3.4in]{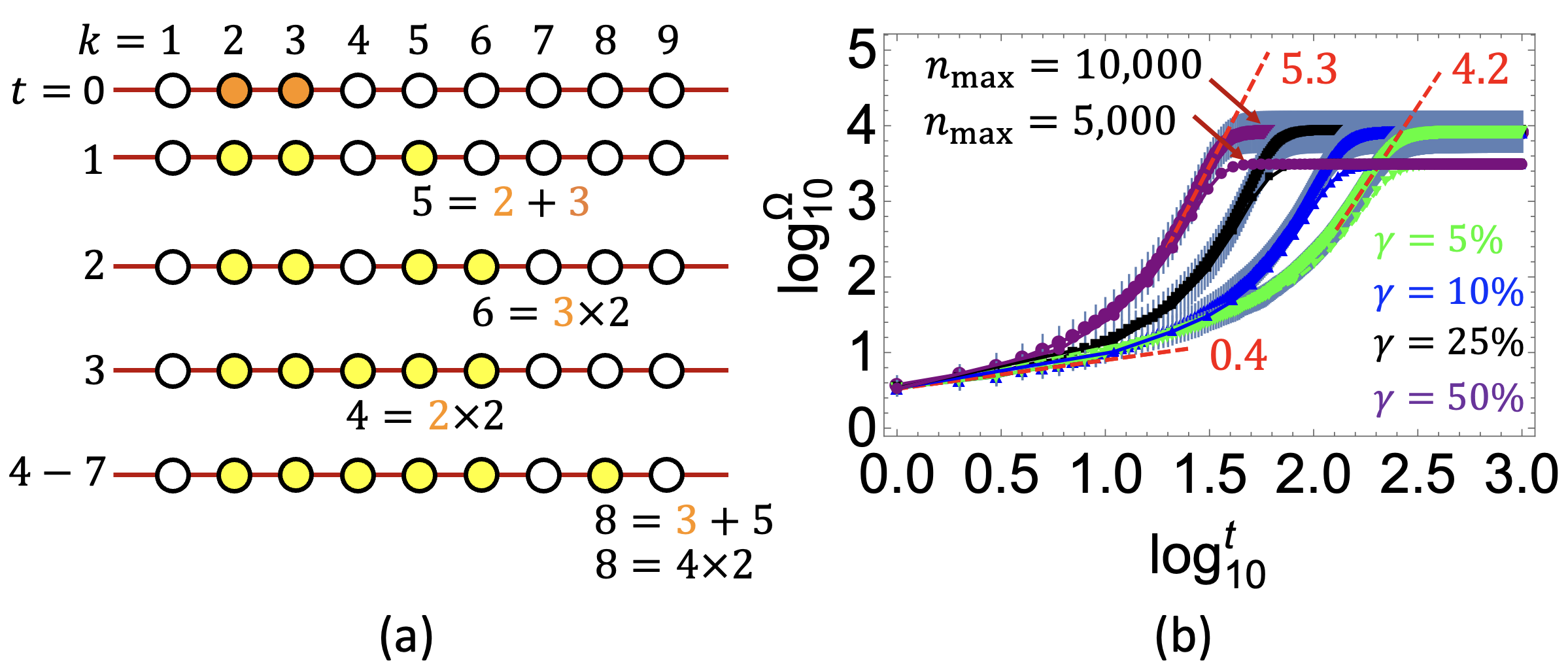}
  \caption{Illustration of the discrete theoretical model for understanding the
  nonlinear nature of frequency proliferation processes. (a) An example for showing the growth
  of the number of frequencies in time. The initial state is characterized by
  the two occupied frequencies (orange dots). The evolution of the state follows
  the prescribed rules; more information is provided in the text. (b) Log-log
  plot of the total number of frequencies $\Omega$ versus time at typical values
  of $n_{\rm{max}}$ and $\gamma$.  $n_{\rm{max}}$ is the total number of
  permissible frequencies, and $\gamma$ is the proportion of the number of
  frequencies involved in the frequency-proliferation process in each time step.
  The slopes of the fitting lines are indicated at the typical sites, showing
  that the discrete model accommodates a range of slope values covering both the
  nonlinear regimes below and above the linear law. The statistical analysis is
  based on 20 independent simulations; the two initial frequencies are randomly
  specified in each simulation. The standard deviations, which are similar for
  both cases of $n_{\rm{max}}=10,000$ and $n_{\rm{max}}=5,000$, are indicated by
  the error bars only for the case of $n_{\rm{max}}=10,000$ for the sake of
  visual clarity.  }
\label{theory}
\end{figure}

In preceding subsection, we discuss the complex dynamics of thermalization
process in terms of the proliferation of frequencies. Figure~\ref{frequency}(d)
shows that the growth of the total number of frequencies exhibits statistical
regularity in the form of the power law. A key finding is that
$\Omega(t)$ is generally a nonlinear function of time, as indicated by the
deviation of the exponent $\alpha$ from unity over a wide range of $v_0$ in both
Figs.~\ref{frequency}(e) and \ref{frequency}(f). The value of $\alpha$ is
dominated by the frequency-mixing processes complicated by the frequency-drift
effect, all of which are affected by the strength of disturbance $v_0$ and the
dimension of lattice motion according to Figs.~\ref{frequency}(e) and
\ref{frequency}(f).  It is a challenge to theoretically derive the $\alpha(v_0)$
curve obtained by numerical experiment.  Here, instead of predicting the
specific value of $\alpha$ at given $v_0$, we propose a simplified discrete
model, focusing on understanding the nonlinearity of the $\Omega(t)$ curve by
incorporating the essential frequency-mixing processes.

For the sake of simplicity, the frequency takes the discrete values: $\omega_i
= i \omega_0$, where $i = 0,1,2,3...n_{\rm{max}}$, and $\omega_0$ is the unit of
the frequency. The time axis is also discretized. The schematic plot of the
model is presented in Fig.~\ref{theory}(a). The initial state is
characterized by a certain number of frequencies. In simulations, two random
frequencies are occupied at $t=0$, as indicated by the orange dots in
Fig.~\ref{theory}(a). In the model, the proliferation of frequencies is driven
by the basic routines of frequency-mixing ($\omega_1\pm \omega_2$),
frequency-doubling ($\omega_1 \rightarrow 2\omega_1$) and frequency-halving
($\omega_1 \rightarrow \omega_1/2$).

Specifically, from $t_j$ to $t_{j+1}$, the number of frequencies is updated by
the following procedure: (1) Randomly select $m/2$ distinct pairs of frequencies
from the $N_j$ occupied frequencies at previous time $t_j$. The even number $m$
is the ceiling of $\gamma N_j$ (minus one if $\lceil \gamma N_j \rceil$ is odd).
$\gamma<1$. The parameter $\gamma$ is the percentage of frequencies
participating in the frequency proliferation process. For each pair of selected
frequencies $\omega_a$ and $\omega_b$, generate a new frequency $\omega=\omega_a
+ \omega_b$ (if $\omega_a + \omega_b \leq n_{\rm{max}}\omega_0$) or
$\omega=|\omega_a - \omega_b|$ with equal probability. (2) Furthermore, randomly
select $\lceil \gamma N_j \rceil$ frequencies from the $N_j$ occupied
frequencies at previous time $t_j$. For each selected frequency $\omega_a$,
generate a new frequency $\omega=2\omega_a$ (if $2\omega_a \leq
n_{\rm{max}}\omega_0$) or $\omega=\omega_a/2$ (if $\omega_a/\omega_0$ is an even
number) with equal probability.

An example of the proliferation of frequencies is depicted in
Fig.~\ref{theory}(a), where the yellow dots represent the occupied frequencies.
At $t=1$, the new frequency of $5$ is generated as the sum of the initial two
frequencies (via the prescribed frequency-mixing mechanism). At $t=2$ and
$t=3$, the new frequencies of $6$ and $4$ arise via the frequency-doubling
mechanism.  From $t=4$ to $t=7$, the generated frequencies coincide with the old
ones, and thus they do not contribute to the increase of $\Omega$. This example
demonstrates how the prescribed rules can accelerate or decelerate the
frequency proliferation process, providing the mechanism for the nonlinear
growth of $\Omega$.

In Fig.~\ref{theory}(b), we show the log-log plot of the total number of
frequencies versus time at typical values of $n_{\rm{max}}$ and $\gamma$.  The
statistical analysis is based on 20 independent simulations starting from
two random initial frequencies. The simulation results converge to the
corresponding curves in Fig.~\ref{theory}(b).  Reducing the value of
$n_{\rm{max}}$ from $10,000$ to $5,000$ primarily affects the saturation value
of $\Omega$; the remaining segments of the curves are almost the same. The
slopes of the fitting lines are indicated at the typical sites. Note that the
slope in a log-log plot is unaffected by the choice of time scale. An important
observation is that the discrete model accommodates a range of slope values
from $0.4$ to $5.3$, covering both the nonlinear regimes below and above the
linear law.

The discrete frequency-mixing model provides a framework for
understanding the nonlinearity of $\Omega(t)$ curves, characterized by non-unity
$\alpha$ values. To further predict how the value of $\alpha$ is deviated from
unity with the variation of $v_0$, relevant detailed dynamical information
shall be put into the model, such as the temporally-varying $\gamma$ and the
frequency-drift effect.

\begin{figure}[t]  
\centering 
\includegraphics[width=3.47in]{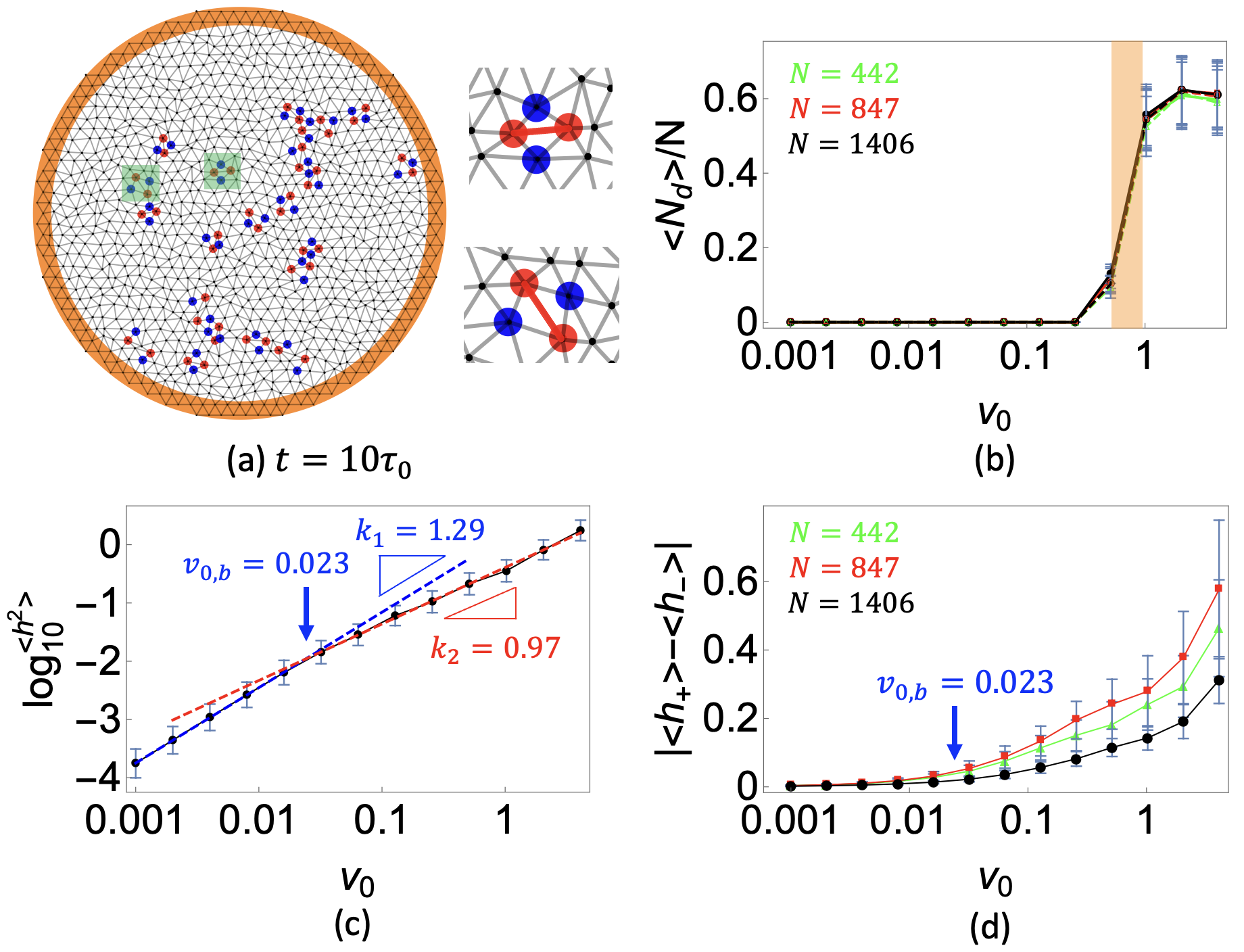}
  \caption{Characterization of the in-plane and out-of-plane fluctuations. (a)
  An instantaneous particle configuration containing topological defects;
  $v_0=0.512$. The red and blue dots represent five- and seven-fold
  disclinations. The insets provide a zoomed-in view of the quadrupoles
  highlighted in green. (b) $\la N_d\ra/N$ is the time-averaged total number of
  disclinations, rescaled by the total number of particles. The data for both
  cases, with and without out-of-plane deformations, collapse on the same curve
  at typical system sizes. The arrow indicates the critical value of $v_0$,
  above which rapid proliferation of topological defects occurs.  (c) Plot of
  the logarithm of $\la h^2\ra$ (the mean squared transverse displacement)
  versus $v_0$.  The values of the slopes $k_1$ and $k_2$ of the fitting lines
  are shown. The cross point of the two fitting lines is located at $v_0 \approx
  0.025$. $N=847$.  (d) Plot of $|\la h_{+}\ra - \la h_{-}\ra|$ versus $v_0$
  shows the breaking of the up-down symmetry with the increase of $v_0$.  $\la
  h_{+}\ra$ and $\la h_{-}\ra$ are the magnitude of the transverse deformation
  along the $z$ and -$z$ directions, respectively.  The critical value of $v_0$,
  above which the up-down symmetry is appreciably broken, coincides with the
  cross point in (c).  The statistical analysis in (b)-(d) is based on the
  sampling during $t\in [\tau_0, 100\tau_0]$ at the resolution of $\Delta
  t=2\tau_0$. The standard deviations are indicated by the error bars.  }
\label{deformation}
\end{figure}

\subsection{Analysis of in-plane and out-of-plane fluctuations}

We proceed to discuss the persistent fluctuations of the drum system as a
dissipationless Hamiltonian system upon the initial disturbance. In this
subsection, the temporally-varying particle configurations are systematically
analyzed from both perspectives of in-plane and out-of-plane deformations.

The Delaunay triangulation procedure serves as a suitable tool to characterize
the in-plane deformation of the lattice~\cite{nelson2002defects}. First, we
project all particles in the deformed lattice onto the $x-y$ plane by setting
their $z$-coordinates to zero. For a collection of particles on the plane,
geometric bonds can be uniquely established among adjacent particles based on
the principle of maximizing the minimum angle in the triangulated configuration.
Note that the geometric bonds established by the Delaunay triangulation
algorithm can be distinct from the physical bonds (springs) connecting
particles. In the resulting triangulated particle configuration, topological
defects can be identified, providing information on the relative positions of
particles and the degree of
inhomogeneity~\cite{wojciechowski1996minimum,Mughal2007,Mughal2009,Yao2013a,Soni2018,silva2020formation,yao2024intrinsic}.
In a triangular lattice, the elementary topological defects are $n$-fold
disclinations, referring to particles surrounded by $n$ nearest neighbors with
$n \neq 6$~\cite{nelson2002defects}. A pair of five- and seven-fold
disclinations constitute a dislocation, analogous to an electric dipole.
In connection to mechanical deformation, the concept of dislocation
can be used to characterize the deformation whose effect is to insert an array
of particles into a triangular lattice~\cite{Landau1986}.

In Fig.~\ref{deformation}(a), we show a typical instantaneous particle
configuration projected onto the plane, where the five- and seven-fold
disclinations are represented by red and blue dots.  Besides isolated five- and
seven-disclination pairs (dislocations), we also find isolated quadrupoles
consisting of four disclinations of opposite signs organized in a square
configuration; two of them are highlighted in green and zoomed in for
closer inspection. Here, we discuss the connection of the quadrupole structure
and mechanical deformation. A quadrupole is formed by a local stretching of the
red bond connecting the pair of five-fold disclinations. Specifically, when the
angle between the diagonal (the red line) and side (the black line) bonds of the
quadrupole's square configuration reduces from $60$ degrees (perfect triangular
lattice) to less than $45$ degrees, the original geometric bond between the two
red dots is flipped to connect the blue dots according to the rule of Delaunay
triangulation~\cite{nelson2002defects}.  The flip of the geometric bond converts
the four particles of coordination number six in the perfect triangular lattice
into five- and seven-fold disclinations. The quadrupole defect thus serves as a
particularly effective indicator of local thermal fluctuations in the strain
field. Under sufficiently strong thermal fluctuations, the unbinding of
quadrupoles leads to isolated dislocations and isolated disclinations; more
information is provided in Appendix B. The consecutive unbinding of compound
topological defects at increasing temperature constitutes the classical scenario
of 2D crystal
melting~\cite{halperin1978theory,strandburg1988two,nelson2002defects}.

Figure~\ref{deformation}(b) shows the $v_0$-dependence of the ratio of the
average number of disclinations to the total number of particles.
Within the range of $v_0\in [0.5, 1.0]$ indicated by the orange bar,
we observe the rapid proliferation of topological defects as characterized by
the rapid rise of the $\la N_d\ra/N$ curve.  Here, we notice the concurrence
of the rapid rise of both the $\alpha(v_0)$ curve [in Figs.~\ref{frequency}(e)
and \ref{frequency}(f)] and the $\la N_d\ra/N$ curve [in
Fig.~\ref{deformation}(b)]. This key observation implies the correlation of the
rapid proliferations of frequencies (as characterized by the exponent $\alpha$)
and topological defects in the thermalization process.  Similar correlation also
exists in the nonlinear case; more information is provided in Appendix A.

With the increase of $v_0$, Fig.~\ref{deformation}(b) shows that the
proliferation of topological defects leads to the destruction of crystalline
order, the randomization of particle positions, and, ultimately, thermalization
in coordinate space.  Note that in Fig.~\ref{deformation}(b) the range of the
plot is up to $v_0 = 4.096$. Above this value, some particles are located
outside the circular boundary, invalidating the proper implementation of the
Delaunay triangulation.

Here, the preserved crystalline order in the regime of small $v_0$ is in
contrast to the Hohenberg-Mermin-Wagner theorems, which preclude the
spontaneous breaking of a continuous symmetry in infinite systems with
sufficiently short-range interactions in dimensions $d\leq
2$~\cite{hohenberg1967existence,MW1966,coleman1973there}. The suppression of
topological defects and the preservation of the crystalline order in the drum
system may be attributed to the following reasons: (1) the finite-size effect;
(2) the limited simulation time; (3) the presence of out-of-plane fluctuations.
To check the possibility of the proliferation of topological defects in larger
systems, we perform simulations with $N=\{3865, 6181, 9055, 14386\}$. Even with
these larger systems, we still observe the preserved crystalline order for small
$v_0$ ($v_0 = \{0.001, 0.01, 0.1, 0.2\}$) up to at least $t=100\tau_0$.  In the
ensuing paragraphs, we discuss the latter two effects.

We extend the simulation time to $t=10,000\tau_0$, focusing on whether
topological defect may proliferate under small $v_0$.  Note that the simulation time
for the cases in Fig.~\ref{deformation}(b) is until $t = 100\tau_0$. 
During the ten million simulation steps at $h=10^{-3}$, the energy is well
conserved; the relative variation of the total energy is at the order $10^{-4}$
as the value of $v_0$ is varied from $0.001$ to $0.5$. It turns out that the
defect-free states under small $v_0$ ($v_0 = \{0.001, 0.01, 0.1, 0.2\}$) persist
for extended simulation time up to $t=10,000\tau_0$ (ten million simulation
steps). At $v_0 =0.3$, frame-by-frame examination of 50 evenly sampled
instantaneous particle configurations during $t\in [\tau_0, 10,000\tau_0]$
reveals that most configurations are defect free, with the following exceptions:
a single isolated quadrupole in 10 configurations, 3 isolated quadrupoles in one
configuration, and a pair of anti-parallel dislocations separated by two lattice
spacings in one configuration. These observations are consistent with the
results shown in Fig.~\ref{deformation}(b) obtained at the shorter simulation
time.  Here, we note that due to the limitations of computational approach, the
possibility for the emergence of topological defects in even longer time cannot
be definitely ruled out; the accumulated computational error may obscure the
true long-term dynamic behavior.

Could the preserved crystalline order be related to the out-of-plane
fluctuation? To address this question, we also perform long-time simulations in
the absence of out-of-plane deformation up to $t=10,000\tau_0$. The motion of
the particles in the drum system is confined on the plane in these simulations.
The results are similar to the case allowing out-of-plane deformation as
discussed in the preceding paragraph. Specifically, the lattice is defect free
for $v_0 = \{0.001, 0.01, 0.1, 0.2\}$. At $v_0 =0.3$, we find the presence of a
single isolated quadrupole in 6 out of 50 evenly sampled instantaneous
particle configurations during $t\in [\tau_0, 10,000\tau_0]$; the remaining
configurations are defect free. To conclude, the crystalline order in the regime
of small $v_0$ is still well preserved in long-time simulations (up to
$t=10,000\tau_0$) even in the absence of the out-of-plane fluctuation.

For the out-of-plane deformation, we analyze the mean squared transverse
displacement (along the $z$ axis), and reveal another transition point $v_{0,b}$
as shown in Fig.~\ref{deformation}(c). Specifically, in the log-log plot of $\la
h^2\ra(v_0)$ versus $v_0$, we observe the bending of the fitting line at
$v_0=v_{0,b}$, indicating that the $\la h^2\ra(v_0)$ curve conforms to distinct
power laws across the transition point $v_{0,b}$. In other words, the
out-of-plane fluctuations exhibit two-stage behaviors with the increase of the
disturbance strength. The slope of the fitting line is appreciably reduced from
$k_1=1.29$ to $k_2=0.97$ across $v_{0,b}$. The value of $v_{0,b}$
($v_{0,b}=0.023$) is much smaller than that of the critical $v_0$ above which we
observe the concurrent rapid proliferations of frequencies and topological
defects.

Examination of systems of varying size ($N \in \{442, 847, 1406, 2077, 2912\}$)
shows that the exponents $k_1$ and $k_2$ in the power law governing $\la
h^2\ra(v_0)$ are insensitive to a more than six-fold increase in $N$.
Specifically, $k_1 = 1.21  \pm 0.06$, $k_2 = 0.99 \pm 0.02$, $v_{0,b} = 0.023
\pm 0.001$, and $\sqrt{\la h^2\ra}=0.091 \pm 0.009$.  According to
Eq.(\ref{T_v0}), $T\propto v_0^2$.  Therefore, we obtain the scaling relation of
the mean squared transverse displacement $\la h^2\ra$ with $v_0$:
\be
\la h^2 \ra \propto T^{\beta},
\ee
where $\beta=k_1/2$ and $k_2/2$ in the two regimes across $v_{0,b}$.  The value
of $\beta$ is smaller than unity in both regimes. As such, the anchored drum
system exhibits distinct out-of-plane fluctuation behaviors compared with an
infinitely large elastic membrane in thermal equilibrium.  In the latter system,
the mean squared displacement is proportional to temperature~\cite{Nelson2004c}.
In the drum system, the anchored boundary condition leads to fractional values
for the exponent in the power laws governing the out-of-plane fluctuations.

The pronounced bending of the line in Fig.~\ref{deformation}(c), observed across
different system sizes, leads us to explore the physical origin of this
phenomenon. At $v_0=v_{0,b}$, the system remains free of topological defects.
Spectral analysis also shows that no abrupt behavior occurs in the rate of the
proliferation of frequencies across $v_{0,b}$ along the $v_0$ axis.

We try to understand the transition point at $v_0=v_{0,b}$ from the viewpoint of
symmetry breaking. Specifically, can the bending of the line in
Fig.~\ref{deformation}(c) be related to the breaking of the up-down symmetry in
the out-of-plane fluctuations that may occur at a small value of $v_0$? To
check this possibility, we examine the variation of $\la h_{+}\ra - \la
h_{-}\ra$ with the increase of $v_0$. $h_{+}$ and $h_{-}$ are the magnitude of
the transverse deformation along the $z$ and -$z$ directions, respectively.
The mean values of $h_{+}$ and $h_{-}$ are: $\la h_{+}\ra = \sum_{i;
z_i>0} |z_{i}|$, and $\la h_{-}\ra = \sum_{i; z_i<0} |z_{i}|$, where $z_i$ is the
$z$ coordinate of particle $i$. The summation is over all particles with $z_i>0$
and $z_i<0$, respectively.  The degree of the up-down symmetry breaking is
measured by the quantity $\la h_{+}\ra - \la h_{-}\ra$. The plot of
$\la h_{+}\ra - \la h_{-}\ra$ against $v_0$ is shown in
Fig.~\ref{deformation}(d). We see that $\la h_{+}\ra - \la h_{-}\ra$ gradually
increases with $v_0$. There is no transition point at which the value of $\la
h_{+}\ra - \la h_{-}\ra$ is abruptly deviation from zero. However, the value of
$\la h_{+}\ra - \la h_{-}\ra$ is observed to be appreciably deviated from zero
at the indicated point $v_0=v_{0,b}$, the transition point of the curve in
Fig.~\ref{deformation}(c).  Specifically, for the cases of $N=442$, $847$, and
$1406$, $(\la h_{+}\ra - \la h_{-}\ra)/\ell_0 = 3.5\%$, $4.0\%$, $1.8\%$ at
$v_0=v_{0,b}$.  In other words, when the up-down symmetry breaking, quantified
by $\la h_{+}\ra - \la h_{-}\ra$, exceeds a threshold, the system's
susceptibility ($\partial \log(\la h^2\ra)/\partial v_0$) exhibits an abrupt
change as shown in Fig.~\ref{deformation}(c).  Here, the revealed connection
between the two-stage fluctuation behaviors [Fig.~\ref{deformation}(c)] and the
broken up-down symmetry may provide an important clue for fully understanding
the transition of the system's susceptibility.

\section{Conclusion}

In summary, based on the harmonic lattice model, we investigated the
thermalization process in both velocity and coordinate spaces from the multiple
perspectives of velocity relaxation, spectral structure, and particle position
fluctuations. Specifically, we show the significantly faster relaxation of the
longitudinal component of the velocity compared with that of the transverse
component. Through spectral analysis of the fluctuating kinetic energy curve, we
identify the power law that governs the nonlinear proliferation of
frequencies, revealing a statistical regularity in the complex dynamic evolution
of the particle configuration. The rapid proliferation of frequencies is
observed to coincide with the emergence of topological defects at the same
critical disturbance strength.  We further show that the drum system's
persistent out-of-plane deformations exhibit two-stage fluctuation behaviors,
characterized by distinct power laws and associated with the broken up-down
symmetry. These results advance our understanding of the fundamental
thermalization phenomenon through an examination of microscopic dynamics on the
atomic level. The harmonic drum model can be generalized by incorporating the
element of dissipation for modeling the interaction of the system with the
environment. It is of interest to explore the impact of viscosity on the
spectral structure, fluctuation behavior, and, especially, the decay of dynamic
modes. 
\\

 \noindent {\bf{Acknowledgements}}
This work was supported by the National Natural Science Foundation of China
(Grants No. BC4190050).

\section*{Appendix A: Anharmonic interaction}

\begin{figure}[t]  
\centering 
\includegraphics[width=3.5in]{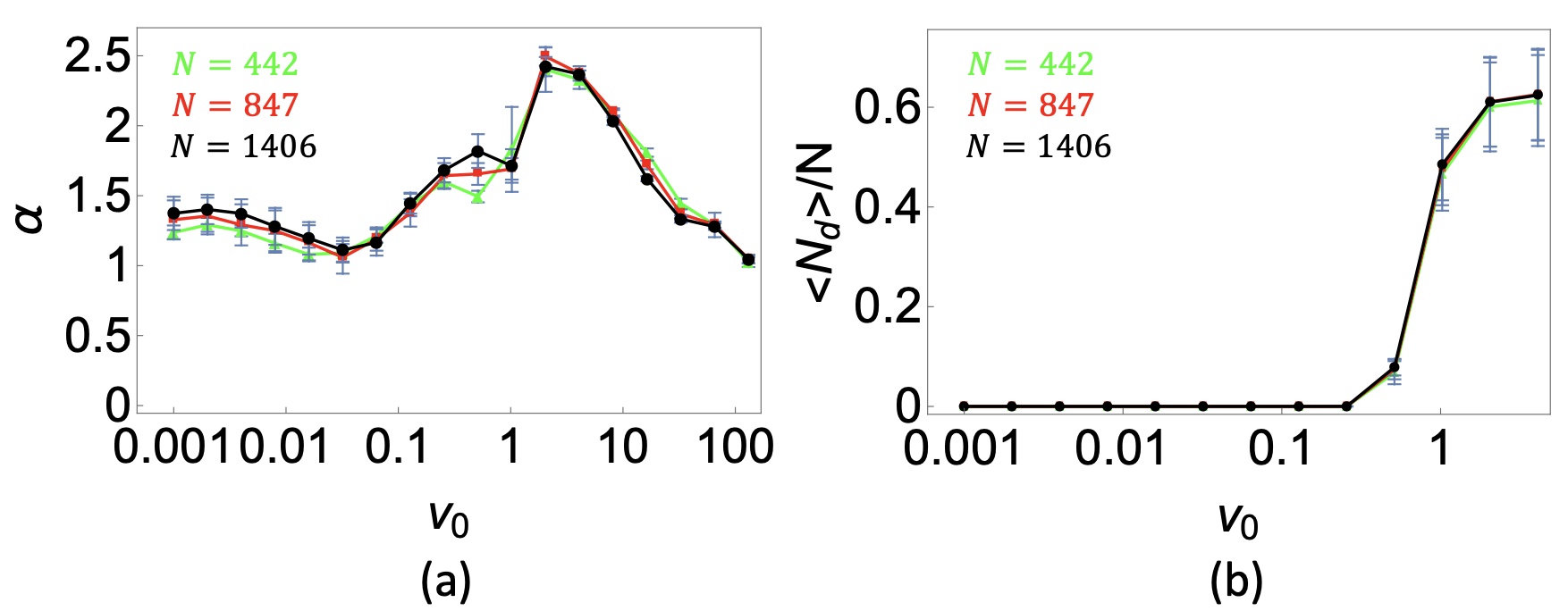}
  \caption{Proliferation of frequencies and topological defects under the
  anharmonic interaction. $\alpha_3=\alpha_4= 1$ (see the text for more
  information). (a) Plot of the exponent $\alpha$ in the power law of
  $\Omega(t)$ versus $v_0$ at typical system sizes.  The statistical analysis
  is based on 10 independent initial distributions of velocity. The sampling is
  during $t\in [\tau_0, 100\tau_0]$ at the resolution of $\Delta t=0.01\tau_0$.
  (b) Plot of $\la N_d\ra/N$ versus $v_0$ at typical system sizes.  $\la
  N_d\ra/N$ is the ratio of the time-averaged total number of disclinations and
  the total number of particles.  The data corresponding to system sizes of $N
  \in \{442, 847, 1406\}$ collapse on the same curve. The arrows indicate the
  same critical value of $v_0$, above which rapid proliferations of frequencies
  and topological defects occur.   
  }
\label{nonlinear}
\end{figure}

\begin{figure*}[t]  
\centering 
\includegraphics[width=6in]{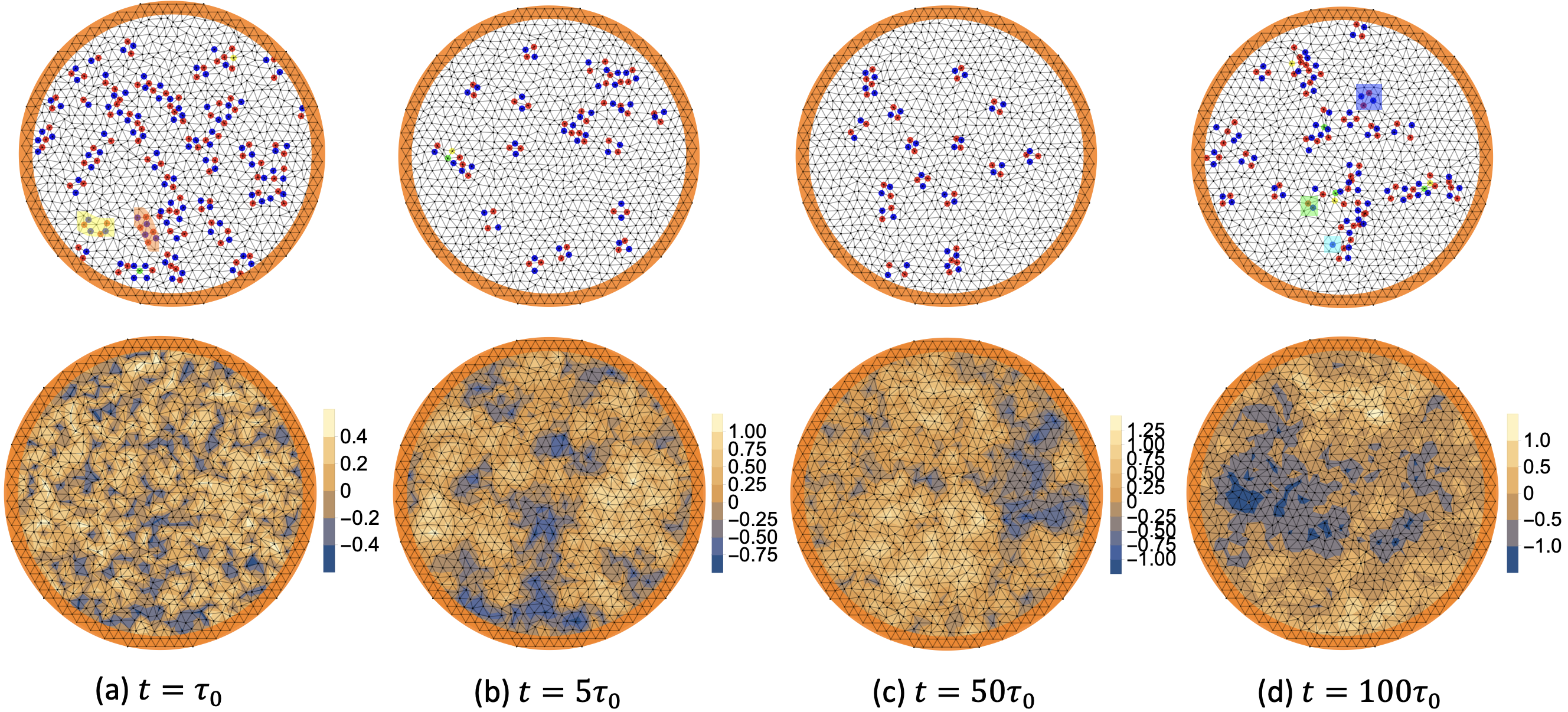}
  \caption{Evolution of the particle configuration characterized by the in-plane
  (upper panels) and out-of-plane (lower panels) deformations. The four-, five-,
  seven- and eight-fold disclinations are colored in green, red, blue, and
  yellow, respectively. $v_0=0.512$. $N = 847$.   
  }
\label{app_B}
\end{figure*}

In this Appendix, we present the main results on the proliferations of
frequencies and topological defects under the anharmonic interaction:
\be
V(r_{\alpha}) = V_0(r_{\alpha}) + \f{1}{3}\alpha_3
(r_{\alpha}-\ell_0)^3+\f{1}{4}\alpha_4 (r_{\alpha}-\ell_0)^4,
\ee
where $r_{\alpha}$ and $\ell_0$ are the actual and rest length of the nonlinear
spring $\alpha$. $V_0(r_{\alpha})$ is the harmonic potential given in
Eq.(\ref{harmonic}). Here, we consider the anharmonic potential up to the fourth
order.

In the presence of the nonlinear terms in the interaction potential, we observe
that the proliferation of frequencies also conforms to the power law as in the
harmonic case. The dependence of the exponent $\alpha$ on $v_0$ is given in
Fig.~\ref{nonlinear}(a). The undulating $\alpha(v_0)$ curve is similar to the
curve of the harmonic case in Fig.~\ref{frequency}(e).  In
Fig.~\ref{nonlinear}(b), we show the variation of the rescaled average number of
disclinations with the increase of $v_0$ at typical system sizes.  These curves
closely resemble those of the harmonic case in Fig.~\ref{deformation}(b).

\section*{Appendix B: Temporally-varying deformations of the lattice}

In this Appendix, we present typical instantaneous particle configurations in
the persistent deformations of the drum system upon the initial disturbance.

The in-plane deformation is characterized by the emergent topological defects;
see the upper panels in Fig.~\ref{app_B}. The four-, five-, seven- and
eight-fold disclinations are represented by the colored dots in green, red,
blue, and yellow, respectively. Typical kinds of defect motifs are highlighted
in the upper panels in Fig.~\ref{app_B}(a) and \ref{app_B}(d), including
quadrupole pile (orange), pleat (yellow), isolated quadrupole (blue),
dislocation (green) and disclination (cyan).

In the out-of-plane deformation, the $z$ axis (height) of each particle is
represented in the contour plots in the lower panels in Fig.~\ref{app_B}. The
highly irregular out-of-plane fluctuations, characterized by the
temporally-varying peak (brighter regions) and valley (darker regions)
structures, are dominated by the growing number of frequencies in time. The
in-plane and out-of-plane deformations, as shown in the upper and lower panels
in Fig.~\ref{app_B}, appear uncorrelated.


\begin{thebibliography}{70}%
\makeatletter
\providecommand \@ifxundefined [1]{%
 \@ifx{#1\undefined}
}%
\providecommand \@ifnum [1]{%
 \ifnum #1\expandafter \@firstoftwo
 \else \expandafter \@secondoftwo
 \fi
}%
\providecommand \@ifx [1]{%
 \ifx #1\expandafter \@firstoftwo
 \else \expandafter \@secondoftwo
 \fi
}%
\providecommand \natexlab [1]{#1}%
\providecommand \enquote  [1]{``#1''}%
\providecommand \bibnamefont  [1]{#1}%
\providecommand \bibfnamefont [1]{#1}%
\providecommand \citenamefont [1]{#1}%
\providecommand \href@noop [0]{\@secondoftwo}%
\providecommand \href [0]{\begingroup \@sanitize@url \@href}%
\providecommand \@href[1]{\@@startlink{#1}\@@href}%
\providecommand \@@href[1]{\endgroup#1\@@endlink}%
\providecommand \@sanitize@url [0]{\catcode `\\12\catcode `\$12\catcode
  `\&12\catcode `\#12\catcode `\^12\catcode `\_12\catcode `\%12\relax}%
\providecommand \@@startlink[1]{}%
\providecommand \@@endlink[0]{}%
\providecommand \url  [0]{\begingroup\@sanitize@url \@url }%
\providecommand \@url [1]{\endgroup\@href {#1}{\urlprefix }}%
\providecommand \urlprefix  [0]{URL }%
\providecommand \Eprint [0]{\href }%
\providecommand \doibase [0]{https://doi.org/}%
\providecommand \selectlanguage [0]{\@gobble}%
\providecommand \bibinfo  [0]{\@secondoftwo}%
\providecommand \bibfield  [0]{\@secondoftwo}%
\providecommand \translation [1]{[#1]}%
\providecommand \BibitemOpen [0]{}%
\providecommand \bibitemStop [0]{}%
\providecommand \bibitemNoStop [0]{.\EOS\space}%
\providecommand \EOS [0]{\spacefactor3000\relax}%
\providecommand \BibitemShut  [1]{\csname bibitem#1\endcsname}%
\let\auto@bib@innerbib\@empty
\bibitem [{\citenamefont {Gibbs}(2014)}]{gibbs2014elementary}%
  \BibitemOpen
  \bibfield  {author} {\bibinfo {author} {\bibfnamefont {J.~W.}\ \bibnamefont
  {Gibbs}},\ }\href@noop {} {\emph {\bibinfo {title} {Elementary Principles in
  Statistical Mechanics}}}\ (\bibinfo  {publisher} {Courier Corporation,
  Massachusetts},\ \bibinfo {year} {2014})\BibitemShut {NoStop}%
\bibitem [{\citenamefont {Khinchin}(1949)}]{aleksandr1949mathematical}%
  \BibitemOpen
  \bibfield  {author} {\bibinfo {author} {\bibfnamefont {A.}~\bibnamefont
  {Khinchin}},\ }\href@noop {} {\emph {\bibinfo {title} {Mathematical
  Foundations of Statistical Mechanics}}}\ (\bibinfo  {publisher} {Courier
  Corporation, Massachusetts},\ \bibinfo {year} {1949})\BibitemShut {NoStop}%
\bibitem [{\citenamefont {Landau}\ and\ \citenamefont
  {Lifshitz}(1999)}]{Landau1999a}%
  \BibitemOpen
  \bibfield  {author} {\bibinfo {author} {\bibfnamefont {L.}~\bibnamefont
  {Landau}}\ and\ \bibinfo {author} {\bibfnamefont {E.}~\bibnamefont
  {Lifshitz}},\ }\href@noop {} {\emph {\bibinfo {title} {Statistical
  Physics}}}\ (\bibinfo  {publisher} {Butterworth-Heinemann, Oxford, UK},\
  \bibinfo {year} {1999})\BibitemShut {NoStop}%
\bibitem [{\citenamefont {Krylov}(2014)}]{krylov2014works}%
  \BibitemOpen
  \bibfield  {author} {\bibinfo {author} {\bibfnamefont {N.~S.}\ \bibnamefont
  {Krylov}},\ }\href@noop {} {\emph {\bibinfo {title} {Works On the Foundations
  of Statistical Physics}}}\ (\bibinfo  {publisher} {Princeton University
  Press, NJ, USA},\ \bibinfo {year} {2014})\BibitemShut {NoStop}%
\bibitem [{\citenamefont {Glansdorff}\ and\ \citenamefont
  {Prigogine}(1971)}]{prigogine1971}%
  \BibitemOpen
  \bibfield  {author} {\bibinfo {author} {\bibfnamefont {P.}~\bibnamefont
  {Glansdorff}}\ and\ \bibinfo {author} {\bibfnamefont {I.}~\bibnamefont
  {Prigogine}},\ }\href@noop {} {\emph {\bibinfo {title} {Thermodynamic Theory
  of Structure, Stability and Fluctuations}}}\ (\bibinfo  {publisher} {John
  Wiley \& Sons Ltd, New Jersey},\ \bibinfo {year} {1971})\BibitemShut
  {NoStop}%
\bibitem [{\citenamefont {Nelson}\ \emph {et~al.}(2004)\citenamefont {Nelson},
  \citenamefont {Piran},\ and\ \citenamefont {Weinberg}}]{Nelson2004c}%
  \BibitemOpen
  \bibfield  {author} {\bibinfo {author} {\bibfnamefont {D.~R.}\ \bibnamefont
  {Nelson}}, \bibinfo {author} {\bibfnamefont {T.}~\bibnamefont {Piran}},\ and\
  \bibinfo {author} {\bibfnamefont {S.}~\bibnamefont {Weinberg}},\ }\href@noop
  {} {\emph {\bibinfo {title} {Statistical Mechanics of Membranes and
  Surfaces}}}\ (\bibinfo  {publisher} {World Scientific, Singapore},\ \bibinfo
  {year} {2004})\BibitemShut {NoStop}%
\bibitem [{\citenamefont {Galliano}\ \emph {et~al.}(2023)\citenamefont
  {Galliano}, \citenamefont {Cates},\ and\ \citenamefont
  {Berthier}}]{galliano2023two}%
  \BibitemOpen
  \bibfield  {author} {\bibinfo {author} {\bibfnamefont {L.}~\bibnamefont
  {Galliano}}, \bibinfo {author} {\bibfnamefont {M.~E.}\ \bibnamefont
  {Cates}},\ and\ \bibinfo {author} {\bibfnamefont {L.}~\bibnamefont
  {Berthier}},\ }\bibfield  {title} {\bibinfo {title} {Two-dimensional crystals
  far from equilibrium},\ }\href@noop {} {\bibfield  {journal} {\bibinfo
  {journal} {Phys. Rev. Lett.}\ }\textbf {\bibinfo {volume} {131}},\ \bibinfo
  {pages} {047101} (\bibinfo {year} {2023})}\BibitemShut {NoStop}%
\bibitem [{\citenamefont {Kosterlitz}\ and\ \citenamefont
  {Thouless}(1973)}]{Kosterlitz1973}%
  \BibitemOpen
  \bibfield  {author} {\bibinfo {author} {\bibfnamefont {J.~M.}\ \bibnamefont
  {Kosterlitz}}\ and\ \bibinfo {author} {\bibfnamefont {D.~J.}\ \bibnamefont
  {Thouless}},\ }\bibfield  {title} {\bibinfo {title} {Ordering, metastability
  and phase transitions in two-dimensional systems},\ }\href@noop {} {\bibfield
   {journal} {\bibinfo  {journal} {J. Phys. C: Solid State Phys.}\ }\textbf
  {\bibinfo {volume} {6}},\ \bibinfo {pages} {1181} (\bibinfo {year}
  {1973})}\BibitemShut {NoStop}%
\bibitem [{\citenamefont {Nishimori}\ and\ \citenamefont
  {Ortiz}(2011)}]{elementsNishi}%
  \BibitemOpen
  \bibfield  {author} {\bibinfo {author} {\bibfnamefont {H.}~\bibnamefont
  {Nishimori}}\ and\ \bibinfo {author} {\bibfnamefont {G.}~\bibnamefont
  {Ortiz}},\ }\href@noop {} {\emph {\bibinfo {title} {Elements of Phase
  Transitions and Critical Phenomena}}}\ (\bibinfo  {publisher} {Oxford
  University Press, Oxford, UK},\ \bibinfo {year} {2011})\BibitemShut {NoStop}%
\bibitem [{\citenamefont {Maxwell}(1860)}]{maxwell1860v}%
  \BibitemOpen
  \bibfield  {author} {\bibinfo {author} {\bibfnamefont {J.~C.}\ \bibnamefont
  {Maxwell}},\ }\bibfield  {title} {\bibinfo {title} {V. illustrations of the
  dynamical theory of gases.—part i. on the motions and collisions of
  perfectly elastic spheres},\ }\href@noop {} {\bibfield  {journal} {\bibinfo
  {journal} {Philos. Mag.}\ }\textbf {\bibinfo {volume} {19}},\ \bibinfo
  {pages} {19} (\bibinfo {year} {1860})}\BibitemShut {NoStop}%
\bibitem [{\citenamefont {Boltzmann}(1964)}]{boltzmann1964lectures}%
  \BibitemOpen
  \bibfield  {author} {\bibinfo {author} {\bibfnamefont {L.}~\bibnamefont
  {Boltzmann}},\ }\href@noop {} {\emph {\bibinfo {title} {Lectures On Gas
  Theory}}}\ (\bibinfo  {publisher} {University of California Press,
  Berkeley},\ \bibinfo {year} {1964})\BibitemShut {NoStop}%
\bibitem [{\citenamefont {Ehrenfest}\ and\ \citenamefont
  {Ehrenfest}(2002)}]{ehrenfest2002conceptual}%
  \BibitemOpen
  \bibfield  {author} {\bibinfo {author} {\bibfnamefont {P.}~\bibnamefont
  {Ehrenfest}}\ and\ \bibinfo {author} {\bibfnamefont {T.}~\bibnamefont
  {Ehrenfest}},\ }\href@noop {} {\emph {\bibinfo {title} {The Conceptual
  Foundations of The Statistical Approach in Mechanics}}}\ (\bibinfo
  {publisher} {Courier Corporation, Massachusetts},\ \bibinfo {year}
  {2002})\BibitemShut {NoStop}%
\bibitem [{\citenamefont {Sinai}(1989)}]{sinai1989dynamical}%
  \BibitemOpen
  \bibfield  {author} {\bibinfo {author} {\bibfnamefont {I.~G.}\ \bibnamefont
  {Sinai}},\ }\href@noop {} {\emph {\bibinfo {title} {Dynamical Systems II:
  Ergodic Theory with Applications to Dynamical Systems and Statistical
  Mechanics}}}\ (\bibinfo  {publisher} {Springer, Berlin},\ \bibinfo {year}
  {1989})\BibitemShut {NoStop}%
\bibitem [{\citenamefont {Prigogine}(2017)}]{prigogine2017non}%
  \BibitemOpen
  \bibfield  {author} {\bibinfo {author} {\bibfnamefont {I.}~\bibnamefont
  {Prigogine}},\ }\href@noop {} {\emph {\bibinfo {title} {Non-equilibrium
  Statistical Mechanics}}}\ (\bibinfo  {publisher} {Dover Publications, New
  York},\ \bibinfo {year} {2017})\BibitemShut {NoStop}%
\bibitem [{\citenamefont {Dumas}(2014)}]{dumas2014kam}%
  \BibitemOpen
  \bibfield  {author} {\bibinfo {author} {\bibfnamefont {H.~S.}\ \bibnamefont
  {Dumas}},\ }\href@noop {} {\emph {\bibinfo {title} {The KAM Story}}}\
  (\bibinfo  {publisher} {World Scientific Publishing Company, Singapore},\
  \bibinfo {year} {2014})\BibitemShut {NoStop}%
\bibitem [{\citenamefont {Callen}\ and\ \citenamefont
  {Welton}(1951)}]{callen1951irreversibility}%
  \BibitemOpen
  \bibfield  {author} {\bibinfo {author} {\bibfnamefont {H.~B.}\ \bibnamefont
  {Callen}}\ and\ \bibinfo {author} {\bibfnamefont {T.~A.}\ \bibnamefont
  {Welton}},\ }\bibfield  {title} {\bibinfo {title} {Irreversibility and
  generalized noise},\ }\href@noop {} {\bibfield  {journal} {\bibinfo
  {journal} {Phys. Rev.}\ }\textbf {\bibinfo {volume} {83}},\ \bibinfo {pages}
  {34} (\bibinfo {year} {1951})}\BibitemShut {NoStop}%
\bibitem [{\citenamefont {Zheng}\ \emph {et~al.}(1996)\citenamefont {Zheng},
  \citenamefont {Hu},\ and\ \citenamefont {Zhang}}]{zheng1996ergodicity}%
  \BibitemOpen
  \bibfield  {author} {\bibinfo {author} {\bibfnamefont {Z.}~\bibnamefont
  {Zheng}}, \bibinfo {author} {\bibfnamefont {G.}~\bibnamefont {Hu}},\ and\
  \bibinfo {author} {\bibfnamefont {J.}~\bibnamefont {Zhang}},\ }\bibfield
  {title} {\bibinfo {title} {Ergodicity in hard-ball systems and boltzmann’s
  entropy},\ }\href@noop {} {\bibfield  {journal} {\bibinfo  {journal} {Phys.
  Rev. E}\ }\textbf {\bibinfo {volume} {53}},\ \bibinfo {pages} {3246}
  (\bibinfo {year} {1996})}\BibitemShut {NoStop}%
\bibitem [{\citenamefont {Frenkel}(2015)}]{frenkel2015order}%
  \BibitemOpen
  \bibfield  {author} {\bibinfo {author} {\bibfnamefont {D.}~\bibnamefont
  {Frenkel}},\ }\bibfield  {title} {\bibinfo {title} {Order through entropy},\
  }\href@noop {} {\bibfield  {journal} {\bibinfo  {journal} {Nat. Mater.}\
  }\textbf {\bibinfo {volume} {14}},\ \bibinfo {pages} {9} (\bibinfo {year}
  {2015})}\BibitemShut {NoStop}%
\bibitem [{\citenamefont {Wang}\ \emph {et~al.}(2024)\citenamefont {Wang},
  \citenamefont {Fu}, \citenamefont {Zhang},\ and\ \citenamefont
  {Zhao}}]{wang2024thermalization}%
  \BibitemOpen
  \bibfield  {author} {\bibinfo {author} {\bibfnamefont {Z.}~\bibnamefont
  {Wang}}, \bibinfo {author} {\bibfnamefont {W.}~\bibnamefont {Fu}}, \bibinfo
  {author} {\bibfnamefont {Y.}~\bibnamefont {Zhang}},\ and\ \bibinfo {author}
  {\bibfnamefont {H.}~\bibnamefont {Zhao}},\ }\bibfield  {title} {\bibinfo
  {title} {Thermalization of two-and three-dimensional classical lattices},\
  }\href@noop {} {\bibfield  {journal} {\bibinfo  {journal} {Phys. Rev. Lett.}\
  }\textbf {\bibinfo {volume} {132}},\ \bibinfo {pages} {217102} (\bibinfo
  {year} {2024})}\BibitemShut {NoStop}%
\bibitem [{\citenamefont {Onsager}(1931)}]{onsager1931reciprocal}%
  \BibitemOpen
  \bibfield  {author} {\bibinfo {author} {\bibfnamefont {L.}~\bibnamefont
  {Onsager}},\ }\bibfield  {title} {\bibinfo {title} {Reciprocal relations in
  irreversible processes. i.},\ }\href@noop {} {\bibfield  {journal} {\bibinfo
  {journal} {Phys. Rev.}\ }\textbf {\bibinfo {volume} {37}},\ \bibinfo {pages}
  {405} (\bibinfo {year} {1931})}\BibitemShut {NoStop}%
\bibitem [{\citenamefont {Kubo}(1966)}]{kubo1966}%
  \BibitemOpen
  \bibfield  {author} {\bibinfo {author} {\bibfnamefont {R.}~\bibnamefont
  {Kubo}},\ }\bibfield  {title} {\bibinfo {title} {The fluctuation-dissipation
  theorem},\ }\href@noop {} {\bibfield  {journal} {\bibinfo  {journal} {Rep.
  Prog. Phys.}\ }\textbf {\bibinfo {volume} {29}},\ \bibinfo {pages} {255}
  (\bibinfo {year} {1966})}\BibitemShut {NoStop}%
\bibitem [{\citenamefont {Li}\ and\ \citenamefont
  {Yorke}(1975)}]{li1975period}%
  \BibitemOpen
  \bibfield  {author} {\bibinfo {author} {\bibfnamefont {T.-Y.}\ \bibnamefont
  {Li}}\ and\ \bibinfo {author} {\bibfnamefont {J.~A.}\ \bibnamefont {Yorke}},\
  }\bibfield  {title} {\bibinfo {title} {Period three implies chaos},\
  }\href@noop {} {\bibfield  {journal} {\bibinfo  {journal} {Am. Math. Mon.}\
  }\textbf {\bibinfo {volume} {82}},\ \bibinfo {pages} {985} (\bibinfo {year}
  {1975})}\BibitemShut {NoStop}%
\bibitem [{\citenamefont {Scheck}(2010)}]{scheck2010mechanics}%
  \BibitemOpen
  \bibfield  {author} {\bibinfo {author} {\bibfnamefont {F.}~\bibnamefont
  {Scheck}},\ }\href@noop {} {\emph {\bibinfo {title} {Mechanics: From Newton's
  Laws to Deterministic Chaos}}}\ (\bibinfo  {publisher} {Springer Science \&
  Business Media, Berlin},\ \bibinfo {year} {2010})\BibitemShut {NoStop}%
\bibitem [{\citenamefont {Kaplan}\ and\ \citenamefont
  {Glass}(2012)}]{kaplan2012understanding}%
  \BibitemOpen
  \bibfield  {author} {\bibinfo {author} {\bibfnamefont {D.}~\bibnamefont
  {Kaplan}}\ and\ \bibinfo {author} {\bibfnamefont {L.}~\bibnamefont {Glass}},\
  }\href@noop {} {\emph {\bibinfo {title} {Understanding Nonlinear Dynamics}}}\
  (\bibinfo  {publisher} {Springer Science \& Business Media, Berlin},\
  \bibinfo {year} {2012})\BibitemShut {NoStop}%
\bibitem [{\citenamefont {Trachenko}\ and\ \citenamefont
  {Brazhkin}(2015)}]{trachenko2015collective}%
  \BibitemOpen
  \bibfield  {author} {\bibinfo {author} {\bibfnamefont {K.}~\bibnamefont
  {Trachenko}}\ and\ \bibinfo {author} {\bibfnamefont {V.}~\bibnamefont
  {Brazhkin}},\ }\bibfield  {title} {\bibinfo {title} {Collective modes and
  thermodynamics of the liquid state},\ }\href@noop {} {\bibfield  {journal}
  {\bibinfo  {journal} {Rep. Prog. Phys.}\ }\textbf {\bibinfo {volume} {79}},\
  \bibinfo {pages} {016502} (\bibinfo {year} {2015})}\BibitemShut {NoStop}%
\bibitem [{\citenamefont {Saporta~Katz}\ and\ \citenamefont
  {Efrati}(2019)}]{PhysRevLett.122.024102}%
  \BibitemOpen
  \bibfield  {author} {\bibinfo {author} {\bibfnamefont {O.}~\bibnamefont
  {Saporta~Katz}}\ and\ \bibinfo {author} {\bibfnamefont {E.}~\bibnamefont
  {Efrati}},\ }\bibfield  {title} {\bibinfo {title} {Self-driven fractional
  rotational diffusion of the harmonic three-mass system},\ }\href@noop {}
  {\bibfield  {journal} {\bibinfo  {journal} {Phys. Rev. Lett.}\ }\textbf
  {\bibinfo {volume} {122}},\ \bibinfo {pages} {024102} (\bibinfo {year}
  {2019})}\BibitemShut {NoStop}%
\bibitem [{\citenamefont {Saporta~Katz}\ and\ \citenamefont
  {Efrati}(2020)}]{PhysRevE.101.032211}%
  \BibitemOpen
  \bibfield  {author} {\bibinfo {author} {\bibfnamefont {O.}~\bibnamefont
  {Saporta~Katz}}\ and\ \bibinfo {author} {\bibfnamefont {E.}~\bibnamefont
  {Efrati}},\ }\bibfield  {title} {\bibinfo {title} {Regular regimes of the
  harmonic three-mass system},\ }\href@noop {} {\bibfield  {journal} {\bibinfo
  {journal} {Phys. Rev. E}\ }\textbf {\bibinfo {volume} {101}},\ \bibinfo
  {pages} {032211} (\bibinfo {year} {2020})}\BibitemShut {NoStop}%
\bibitem [{\citenamefont {Yao}(2022)}]{yao2022collective}%
  \BibitemOpen
  \bibfield  {author} {\bibinfo {author} {\bibfnamefont {Z.}~\bibnamefont
  {Yao}},\ }\bibfield  {title} {\bibinfo {title} {Collective dynamics and
  shattering of disturbed two-dimensional lennard--jones crystals},\
  }\href@noop {} {\bibfield  {journal} {\bibinfo  {journal} {Eur. Phys. J. E}\
  }\textbf {\bibinfo {volume} {45}},\ \bibinfo {pages} {88} (\bibinfo {year}
  {2022})}\BibitemShut {NoStop}%
\bibitem [{\citenamefont {Yao}(2023)}]{yao2023non}%
  \BibitemOpen
  \bibfield  {author} {\bibinfo {author} {\bibfnamefont {Z.}~\bibnamefont
  {Yao}},\ }\bibfield  {title} {\bibinfo {title} {Non-linear dynamics and
  emergent statistical regularity in classical lennard--jones three-body system
  upon disturbance},\ }\href@noop {} {\bibfield  {journal} {\bibinfo  {journal}
  {Eur. Phys. J. B}\ }\textbf {\bibinfo {volume} {96}},\ \bibinfo {pages} {159}
  (\bibinfo {year} {2023})}\BibitemShut {NoStop}%
\bibitem [{\citenamefont {Kadanoff}(1986)}]{kadanoff1986two}%
  \BibitemOpen
  \bibfield  {author} {\bibinfo {author} {\bibfnamefont {L.~P.}\ \bibnamefont
  {Kadanoff}},\ }\bibfield  {title} {\bibinfo {title} {On two levels},\
  }\href@noop {} {\bibfield  {journal} {\bibinfo  {journal} {Physics today}\
  }\textbf {\bibinfo {volume} {39}},\ \bibinfo {pages} {7} (\bibinfo {year}
  {1986})}\BibitemShut {NoStop}%
\bibitem [{\citenamefont {Kadanoff}(1999)}]{Kadanoff1999}%
  \BibitemOpen
  \bibfield  {author} {\bibinfo {author} {\bibfnamefont {L.~P.}\ \bibnamefont
  {Kadanoff}},\ }\href@noop {} {\emph {\bibinfo {title} {From Order to Chaos
  II}}}\ (\bibinfo  {publisher} {World Scientific},\ \bibinfo {year}
  {1999})\BibitemShut {NoStop}%
\bibitem [{\citenamefont {Fermi}\ \emph {et~al.}(1955)\citenamefont {Fermi},
  \citenamefont {Pasta}, \citenamefont {Ulam},\ and\ \citenamefont
  {Tsingou}}]{fermi1955studies}%
  \BibitemOpen
  \bibfield  {author} {\bibinfo {author} {\bibfnamefont {E.}~\bibnamefont
  {Fermi}}, \bibinfo {author} {\bibfnamefont {P.}~\bibnamefont {Pasta}},
  \bibinfo {author} {\bibfnamefont {S.}~\bibnamefont {Ulam}},\ and\ \bibinfo
  {author} {\bibfnamefont {M.}~\bibnamefont {Tsingou}},\ }\href@noop {} {\emph
  {\bibinfo {title} {Studies of the nonlinear problems}}},\ \bibinfo {type}
  {Tech. Rep.}\ (\bibinfo  {institution} {Los Alamos Scientific Lab.},\
  \bibinfo {year} {1955})\BibitemShut {NoStop}%
\bibitem [{\citenamefont {Dauxois}\ \emph {et~al.}(2002)\citenamefont
  {Dauxois}, \citenamefont {Ruffo}, \citenamefont {Arimondo},\ and\
  \citenamefont {Wilkens}}]{dauxois2002dynamics}%
  \BibitemOpen
  \bibfield  {author} {\bibinfo {author} {\bibfnamefont {T.}~\bibnamefont
  {Dauxois}}, \bibinfo {author} {\bibfnamefont {S.}~\bibnamefont {Ruffo}},
  \bibinfo {author} {\bibfnamefont {E.}~\bibnamefont {Arimondo}},\ and\
  \bibinfo {author} {\bibfnamefont {M.}~\bibnamefont {Wilkens}},\ }\href@noop
  {} {\emph {\bibinfo {title} {Dynamics and Thermodynamics of Systems with
  Long-Range Interactions}}}\ (\bibinfo  {publisher} {Springer},\ \bibinfo
  {year} {2002})\BibitemShut {NoStop}%
\bibitem [{\citenamefont {Berman}\ and\ \citenamefont
  {Izrailev}(2005)}]{berman2005fermi}%
  \BibitemOpen
  \bibfield  {author} {\bibinfo {author} {\bibfnamefont {G.}~\bibnamefont
  {Berman}}\ and\ \bibinfo {author} {\bibfnamefont {F.}~\bibnamefont
  {Izrailev}},\ }\bibfield  {title} {\bibinfo {title} {The fermi--pasta--ulam
  problem: fifty years of progress},\ }\href@noop {} {\bibfield  {journal}
  {\bibinfo  {journal} {Chaos}\ }\textbf {\bibinfo {volume} {15}} (\bibinfo
  {year} {2005})}\BibitemShut {NoStop}%
\bibitem [{\citenamefont {Gallavotti}(2007)}]{gallavotti2007fermi}%
  \BibitemOpen
  \bibfield  {author} {\bibinfo {author} {\bibfnamefont {G.}~\bibnamefont
  {Gallavotti}},\ }\href@noop {} {\emph {\bibinfo {title} {The Fermi-Pasta-Ulam
  Problem: A Status Report}}}\ (\bibinfo  {publisher} {Springer Berlin,
  Heidelberg},\ \bibinfo {year} {2007})\BibitemShut {NoStop}%
\bibitem [{\citenamefont {Mulansky}\ \emph {et~al.}(2009)\citenamefont
  {Mulansky}, \citenamefont {Ahnert}, \citenamefont {Pikovsky},\ and\
  \citenamefont {Shepelyansky}}]{mulansky2009dynamical}%
  \BibitemOpen
  \bibfield  {author} {\bibinfo {author} {\bibfnamefont {M.}~\bibnamefont
  {Mulansky}}, \bibinfo {author} {\bibfnamefont {K.}~\bibnamefont {Ahnert}},
  \bibinfo {author} {\bibfnamefont {A.}~\bibnamefont {Pikovsky}},\ and\
  \bibinfo {author} {\bibfnamefont {D.~L.}\ \bibnamefont {Shepelyansky}},\
  }\bibfield  {title} {\bibinfo {title} {Dynamical thermalization of disordered
  nonlinear lattices},\ }\href@noop {} {\bibfield  {journal} {\bibinfo
  {journal} {Phys. Rev. E}\ }\textbf {\bibinfo {volume} {80}},\ \bibinfo
  {pages} {056212} (\bibinfo {year} {2009})}\BibitemShut {NoStop}%
\bibitem [{\citenamefont {Ribeiro-Teixeira}\ \emph {et~al.}(2014)\citenamefont
  {Ribeiro-Teixeira}, \citenamefont {Benetti}, \citenamefont {Pakter},\ and\
  \citenamefont {Levin}}]{ribeiro2014ergodicity}%
  \BibitemOpen
  \bibfield  {author} {\bibinfo {author} {\bibfnamefont {A.~C.}\ \bibnamefont
  {Ribeiro-Teixeira}}, \bibinfo {author} {\bibfnamefont {F.~P.}\ \bibnamefont
  {Benetti}}, \bibinfo {author} {\bibfnamefont {R.}~\bibnamefont {Pakter}},\
  and\ \bibinfo {author} {\bibfnamefont {Y.}~\bibnamefont {Levin}},\ }\bibfield
   {title} {\bibinfo {title} {Ergodicity breaking and quasistationary states in
  systems with long-range interactions},\ }\href@noop {} {\bibfield  {journal}
  {\bibinfo  {journal} {Phys. Rev. E}\ }\textbf {\bibinfo {volume} {89}},\
  \bibinfo {pages} {022130} (\bibinfo {year} {2014})}\BibitemShut {NoStop}%
\bibitem [{\citenamefont {Horn}\ and\ \citenamefont
  {L{\"o}wen}(2014)}]{horn2014does}%
  \BibitemOpen
  \bibfield  {author} {\bibinfo {author} {\bibfnamefont {T.}~\bibnamefont
  {Horn}}\ and\ \bibinfo {author} {\bibfnamefont {H.}~\bibnamefont
  {L{\"o}wen}},\ }\bibfield  {title} {\bibinfo {title} {How does a thermal
  binary crystal break under shear?},\ }\href@noop {} {\bibfield  {journal}
  {\bibinfo  {journal} {J. Chem. Phys.}\ }\textbf {\bibinfo {volume} {141}}
  (\bibinfo {year} {2014})}\BibitemShut {NoStop}%
\bibitem [{\citenamefont {Košmrlj}\ and\ \citenamefont
  {Nelson}(2016)}]{NelsonP2016}%
  \BibitemOpen
  \bibfield  {author} {\bibinfo {author} {\bibfnamefont {A.}~\bibnamefont
  {Košmrlj}}\ and\ \bibinfo {author} {\bibfnamefont {D.~R.}\ \bibnamefont
  {Nelson}},\ }\bibfield  {title} {\bibinfo {title} {Response of thermalized
  ribbons to pulling and bending},\ }\href@noop {} {\bibfield  {journal}
  {\bibinfo  {journal} {Phys. Rev. B}\ }\textbf {\bibinfo {volume} {93}},\
  \bibinfo {pages} {125431} (\bibinfo {year} {2016})}\BibitemShut {NoStop}%
\bibitem [{\citenamefont {Zaccone}(2023)}]{zaccone2023theory}%
  \BibitemOpen
  \bibfield  {author} {\bibinfo {author} {\bibfnamefont {A.}~\bibnamefont
  {Zaccone}},\ }\href@noop {} {\emph {\bibinfo {title} {Theory of Disordered
  Solids}}}\ (\bibinfo  {publisher} {Springer, New York},\ \bibinfo {year}
  {2023})\BibitemShut {NoStop}%
\bibitem [{\citenamefont {Jonay}\ and\ \citenamefont
  {Zhou}(2024)}]{jonay2024physical}%
  \BibitemOpen
  \bibfield  {author} {\bibinfo {author} {\bibfnamefont {C.}~\bibnamefont
  {Jonay}}\ and\ \bibinfo {author} {\bibfnamefont {T.}~\bibnamefont {Zhou}},\
  }\bibfield  {title} {\bibinfo {title} {Physical theory of two-stage
  thermalization},\ }\href@noop {} {\bibfield  {journal} {\bibinfo  {journal}
  {Phys. Rev. B}\ }\textbf {\bibinfo {volume} {110}},\ \bibinfo {pages}
  {L020306} (\bibinfo {year} {2024})}\BibitemShut {NoStop}%
\bibitem [{\citenamefont {Vargas}\ \emph {et~al.}(2025)\citenamefont {Vargas},
  \citenamefont {Puccinelli},\ and\ \citenamefont {Bordin}}]{vargas2025order}%
  \BibitemOpen
  \bibfield  {author} {\bibinfo {author} {\bibfnamefont {A.}~\bibnamefont
  {Vargas}}, \bibinfo {author} {\bibfnamefont {T.}~\bibnamefont {Puccinelli}},\
  and\ \bibinfo {author} {\bibfnamefont {J.~R.}\ \bibnamefont {Bordin}},\
  }\bibfield  {title} {\bibinfo {title} {Order-disorder transitions and thermal
  pathways in frustrated 2d colloidal crystals},\ }\href@noop {} {\bibfield
  {journal} {\bibinfo  {journal} {Braz. J. Phys.}\ }\textbf {\bibinfo {volume}
  {55}},\ \bibinfo {pages} {281} (\bibinfo {year} {2025})}\BibitemShut
  {NoStop}%
\bibitem [{\citenamefont {De~Wijn}\ and\ \citenamefont
  {Fasolino}(2009)}]{de2009relating}%
  \BibitemOpen
  \bibfield  {author} {\bibinfo {author} {\bibfnamefont {A.~S.}\ \bibnamefont
  {De~Wijn}}\ and\ \bibinfo {author} {\bibfnamefont {A.}~\bibnamefont
  {Fasolino}},\ }\bibfield  {title} {\bibinfo {title} {Relating chaos to
  deterministic diffusion of a molecule adsorbed on a surface},\ }\href@noop {}
  {\bibfield  {journal} {\bibinfo  {journal} {J. Phys. Condens. Matter}\
  }\textbf {\bibinfo {volume} {21}},\ \bibinfo {pages} {264002} (\bibinfo
  {year} {2009})}\BibitemShut {NoStop}%
\bibitem [{\citenamefont {Keim}\ \emph {et~al.}(2004)\citenamefont {Keim},
  \citenamefont {Maret}, \citenamefont {Herz},\ and\ \citenamefont {von
  Gr{\"u}nberg}}]{keim2004harmonic}%
  \BibitemOpen
  \bibfield  {author} {\bibinfo {author} {\bibfnamefont {P.}~\bibnamefont
  {Keim}}, \bibinfo {author} {\bibfnamefont {G.}~\bibnamefont {Maret}},
  \bibinfo {author} {\bibfnamefont {U.}~\bibnamefont {Herz}},\ and\ \bibinfo
  {author} {\bibfnamefont {H.-H.}\ \bibnamefont {von Gr{\"u}nberg}},\
  }\bibfield  {title} {\bibinfo {title} {Harmonic lattice behavior of
  two-dimensional colloidal crystals},\ }\href@noop {} {\bibfield  {journal}
  {\bibinfo  {journal} {Phys. Rev. Lett.}\ }\textbf {\bibinfo {volume} {92}},\
  \bibinfo {pages} {215504} (\bibinfo {year} {2004})}\BibitemShut {NoStop}%
\bibitem [{\citenamefont {Landau}\ and\ \citenamefont
  {Lifshitz}(1986)}]{Landau1986}%
  \BibitemOpen
  \bibfield  {author} {\bibinfo {author} {\bibfnamefont {L.~D.}\ \bibnamefont
  {Landau}}\ and\ \bibinfo {author} {\bibfnamefont {E.~M.}\ \bibnamefont
  {Lifshitz}},\ }\href@noop {} {\emph {\bibinfo {title} {Theory of Elasticity,
  3rd edition}}}\ (\bibinfo  {publisher} {Butterworth-Heinemann, Oxford, UK},\
  \bibinfo {year} {1986})\BibitemShut {NoStop}%
\bibitem [{\citenamefont {Audoly}\ and\ \citenamefont
  {Pomeau}(2010)}]{audoly2010elasticity}%
  \BibitemOpen
  \bibfield  {author} {\bibinfo {author} {\bibfnamefont {B.}~\bibnamefont
  {Audoly}}\ and\ \bibinfo {author} {\bibfnamefont {Y.}~\bibnamefont
  {Pomeau}},\ }\href@noop {} {\emph {\bibinfo {title} {Elasticity and
  Geometry}}}\ (\bibinfo  {publisher} {Oxford University Press, Oxford, UK},\
  \bibinfo {year} {2010})\BibitemShut {NoStop}%
\bibitem [{\citenamefont {Halperin}\ and\ \citenamefont
  {Nelson}(1978)}]{halperin1978theory}%
  \BibitemOpen
  \bibfield  {author} {\bibinfo {author} {\bibfnamefont {B.}~\bibnamefont
  {Halperin}}\ and\ \bibinfo {author} {\bibfnamefont {D.~R.}\ \bibnamefont
  {Nelson}},\ }\bibfield  {title} {\bibinfo {title} {Theory of two-dimensional
  melting},\ }\href@noop {} {\bibfield  {journal} {\bibinfo  {journal} {Phys.
  Rev. Lett.}\ }\textbf {\bibinfo {volume} {41}},\ \bibinfo {pages} {121}
  (\bibinfo {year} {1978})}\BibitemShut {NoStop}%
\bibitem [{\citenamefont {Strandburg}(1988)}]{strandburg1988two}%
  \BibitemOpen
  \bibfield  {author} {\bibinfo {author} {\bibfnamefont {K.~J.}\ \bibnamefont
  {Strandburg}},\ }\bibfield  {title} {\bibinfo {title} {Two-dimensional
  melting},\ }\href@noop {} {\bibfield  {journal} {\bibinfo  {journal} {Rev.
  Mod. Phys.}\ }\textbf {\bibinfo {volume} {60}},\ \bibinfo {pages} {161}
  (\bibinfo {year} {1988})}\BibitemShut {NoStop}%
\bibitem [{\citenamefont {Nelson}(2002)}]{nelson2002defects}%
  \BibitemOpen
  \bibfield  {author} {\bibinfo {author} {\bibfnamefont {D.~R.}\ \bibnamefont
  {Nelson}},\ }\href@noop {} {\emph {\bibinfo {title} {Defects and Geometry in
  Condensed Matter Physics}}}\ (\bibinfo  {publisher} {Cambridge University
  Press, Cambridge},\ \bibinfo {year} {2002})\BibitemShut {NoStop}%
\bibitem [{\citenamefont {Blees}\ \emph {et~al.}(2015)\citenamefont {Blees},
  \citenamefont {Barnard}, \citenamefont {Rose}, \citenamefont {Roberts},
  \citenamefont {McGill}, \citenamefont {Huang}, \citenamefont {Ruyack},
  \citenamefont {Kevek}, \citenamefont {Kobrin}, \citenamefont {Muller},\ and\
  \citenamefont {McEuen}}]{blees2015graphene}%
  \BibitemOpen
  \bibfield  {author} {\bibinfo {author} {\bibfnamefont {M.~K.}\ \bibnamefont
  {Blees}}, \bibinfo {author} {\bibfnamefont {A.~W.}\ \bibnamefont {Barnard}},
  \bibinfo {author} {\bibfnamefont {P.~A.}\ \bibnamefont {Rose}}, \bibinfo
  {author} {\bibfnamefont {S.~P.}\ \bibnamefont {Roberts}}, \bibinfo {author}
  {\bibfnamefont {K.~L.}\ \bibnamefont {McGill}}, \bibinfo {author}
  {\bibfnamefont {P.~Y.}\ \bibnamefont {Huang}}, \bibinfo {author}
  {\bibfnamefont {A.~R.}\ \bibnamefont {Ruyack}}, \bibinfo {author}
  {\bibfnamefont {J.~W.}\ \bibnamefont {Kevek}}, \bibinfo {author}
  {\bibfnamefont {B.}~\bibnamefont {Kobrin}}, \bibinfo {author} {\bibfnamefont
  {D.~A.}\ \bibnamefont {Muller}},\ and\ \bibinfo {author} {\bibfnamefont
  {P.~L.}\ \bibnamefont {McEuen}},\ }\bibfield  {title} {\bibinfo {title}
  {Graphene kirigami},\ }\href@noop {} {\bibfield  {journal} {\bibinfo
  {journal} {Nature}\ }\textbf {\bibinfo {volume} {524}},\ \bibinfo {pages}
  {204} (\bibinfo {year} {2015})}\BibitemShut {NoStop}%
\bibitem [{\citenamefont {Wan}\ \emph {et~al.}(2017)\citenamefont {Wan},
  \citenamefont {Nelson},\ and\ \citenamefont {Bowick}}]{wan2017thermal}%
  \BibitemOpen
  \bibfield  {author} {\bibinfo {author} {\bibfnamefont {D.}~\bibnamefont
  {Wan}}, \bibinfo {author} {\bibfnamefont {D.~R.}\ \bibnamefont {Nelson}},\
  and\ \bibinfo {author} {\bibfnamefont {M.~J.}\ \bibnamefont {Bowick}},\
  }\bibfield  {title} {\bibinfo {title} {Thermal stiffening of clamped elastic
  ribbons},\ }\href@noop {} {\bibfield  {journal} {\bibinfo  {journal} {Phys.
  Rev. B}\ }\textbf {\bibinfo {volume} {96}},\ \bibinfo {pages} {014106}
  (\bibinfo {year} {2017})}\BibitemShut {NoStop}%
\bibitem [{\citenamefont {Le~Doussal}\ and\ \citenamefont
  {Radzihovsky}(2021)}]{le2021thermal}%
  \BibitemOpen
  \bibfield  {author} {\bibinfo {author} {\bibfnamefont {P.}~\bibnamefont
  {Le~Doussal}}\ and\ \bibinfo {author} {\bibfnamefont {L.}~\bibnamefont
  {Radzihovsky}},\ }\bibfield  {title} {\bibinfo {title} {Thermal buckling
  transition of crystalline membranes in a field},\ }\href@noop {} {\bibfield
  {journal} {\bibinfo  {journal} {Phys. Rev. Lett.}\ }\textbf {\bibinfo
  {volume} {127}},\ \bibinfo {pages} {015702} (\bibinfo {year}
  {2021})}\BibitemShut {NoStop}%
\bibitem [{\citenamefont {Hanakata}\ \emph {et~al.}(2021)\citenamefont
  {Hanakata}, \citenamefont {Bhabesh}, \citenamefont {Bowick}, \citenamefont
  {Nelson},\ and\ \citenamefont {Yllanes}}]{hanakata2021thermal}%
  \BibitemOpen
  \bibfield  {author} {\bibinfo {author} {\bibfnamefont {P.~Z.}\ \bibnamefont
  {Hanakata}}, \bibinfo {author} {\bibfnamefont {S.~S.}\ \bibnamefont
  {Bhabesh}}, \bibinfo {author} {\bibfnamefont {M.~J.}\ \bibnamefont {Bowick}},
  \bibinfo {author} {\bibfnamefont {D.~R.}\ \bibnamefont {Nelson}},\ and\
  \bibinfo {author} {\bibfnamefont {D.}~\bibnamefont {Yllanes}},\ }\bibfield
  {title} {\bibinfo {title} {Thermal buckling and symmetry breaking in thin
  ribbons under compression},\ }\href@noop {} {\bibfield  {journal} {\bibinfo
  {journal} {Extreme Mech. Lett.}\ }\textbf {\bibinfo {volume} {44}},\ \bibinfo
  {pages} {101270} (\bibinfo {year} {2021})}\BibitemShut {NoStop}%
\bibitem [{\citenamefont {Chen}\ \emph {et~al.}(2022)\citenamefont {Chen},
  \citenamefont {Wan},\ and\ \citenamefont {Bowick}}]{chen2022spontaneous}%
  \BibitemOpen
  \bibfield  {author} {\bibinfo {author} {\bibfnamefont {Z.}~\bibnamefont
  {Chen}}, \bibinfo {author} {\bibfnamefont {D.}~\bibnamefont {Wan}},\ and\
  \bibinfo {author} {\bibfnamefont {M.~J.}\ \bibnamefont {Bowick}},\ }\bibfield
   {title} {\bibinfo {title} {Spontaneous tilt of single-clamped thermal
  elastic sheets},\ }\href@noop {} {\bibfield  {journal} {\bibinfo  {journal}
  {Phys. Rev. Lett.}\ }\textbf {\bibinfo {volume} {128}},\ \bibinfo {pages}
  {028006} (\bibinfo {year} {2022})}\BibitemShut {NoStop}%
\bibitem [{\citenamefont {Rapaport}(2004)}]{rapaport2004art}%
  \BibitemOpen
  \bibfield  {author} {\bibinfo {author} {\bibfnamefont {D.}~\bibnamefont
  {Rapaport}},\ }\href@noop {} {\emph {\bibinfo {title} {The Art of Molecular
  Dynamics Simulation}}}\ (\bibinfo  {publisher} {Cambridge University Press,
  Cambridge, UK},\ \bibinfo {year} {2004})\BibitemShut {NoStop}%
\bibitem [{\citenamefont {Acharya}\ \emph {et~al.}(2020)\citenamefont
  {Acharya}, \citenamefont {Sengupta}, \citenamefont {Chakraborty},\ and\
  \citenamefont {Ramola}}]{acharya2020athermal}%
  \BibitemOpen
  \bibfield  {author} {\bibinfo {author} {\bibfnamefont {P.}~\bibnamefont
  {Acharya}}, \bibinfo {author} {\bibfnamefont {S.}~\bibnamefont {Sengupta}},
  \bibinfo {author} {\bibfnamefont {B.}~\bibnamefont {Chakraborty}},\ and\
  \bibinfo {author} {\bibfnamefont {K.}~\bibnamefont {Ramola}},\ }\bibfield
  {title} {\bibinfo {title} {Athermal fluctuations in disordered crystals},\
  }\href@noop {} {\bibfield  {journal} {\bibinfo  {journal} {Phys. Rev. Lett.}\
  }\textbf {\bibinfo {volume} {124}},\ \bibinfo {pages} {168004} (\bibinfo
  {year} {2020})}\BibitemShut {NoStop}%
\bibitem [{\citenamefont {Maharana}(2022)}]{maharana2022athermal}%
  \BibitemOpen
  \bibfield  {author} {\bibinfo {author} {\bibfnamefont {R.}~\bibnamefont
  {Maharana}},\ }\bibfield  {title} {\bibinfo {title} {Athermal fluctuations in
  three dimensional disordered crystals},\ }\href@noop {} {\bibfield  {journal}
  {\bibinfo  {journal} {J. Stat. Mech. Theory Exp.}\ }\textbf {\bibinfo
  {volume} {2022}},\ \bibinfo {pages} {103201} (\bibinfo {year}
  {2022})}\BibitemShut {NoStop}%
\bibitem [{\citenamefont {Xu}\ \emph {et~al.}(2007)\citenamefont {Xu},
  \citenamefont {Wyart}, \citenamefont {Liu},\ and\ \citenamefont
  {Nagel}}]{xu2007excess}%
  \BibitemOpen
  \bibfield  {author} {\bibinfo {author} {\bibfnamefont {N.}~\bibnamefont
  {Xu}}, \bibinfo {author} {\bibfnamefont {M.}~\bibnamefont {Wyart}}, \bibinfo
  {author} {\bibfnamefont {A.~J.}\ \bibnamefont {Liu}},\ and\ \bibinfo {author}
  {\bibfnamefont {S.~R.}\ \bibnamefont {Nagel}},\ }\bibfield  {title} {\bibinfo
  {title} {Excess vibrational modes and the boson peak in model glasses},\
  }\href@noop {} {\bibfield  {journal} {\bibinfo  {journal} {Phys. Rev. Lett.}\
  }\textbf {\bibinfo {volume} {98}},\ \bibinfo {pages} {175502} (\bibinfo
  {year} {2007})}\BibitemShut {NoStop}%
\bibitem [{\citenamefont {Manning}\ and\ \citenamefont
  {Liu}(2011)}]{manning2011vibrational}%
  \BibitemOpen
  \bibfield  {author} {\bibinfo {author} {\bibfnamefont {M.~L.}\ \bibnamefont
  {Manning}}\ and\ \bibinfo {author} {\bibfnamefont {A.~J.}\ \bibnamefont
  {Liu}},\ }\bibfield  {title} {\bibinfo {title} {Vibrational modes identify
  soft spots in a sheared disordered packing},\ }\href@noop {} {\bibfield
  {journal} {\bibinfo  {journal} {Phys. Rev. Lett.}\ }\textbf {\bibinfo
  {volume} {107}},\ \bibinfo {pages} {108302} (\bibinfo {year}
  {2011})}\BibitemShut {NoStop}%
\bibitem [{\citenamefont {Wu}\ \emph {et~al.}(2023)\citenamefont {Wu},
  \citenamefont {Chen}, \citenamefont {Wang}, \citenamefont {Kob},\ and\
  \citenamefont {Xu}}]{wu2023topology}%
  \BibitemOpen
  \bibfield  {author} {\bibinfo {author} {\bibfnamefont {Z.~W.}\ \bibnamefont
  {Wu}}, \bibinfo {author} {\bibfnamefont {Y.}~\bibnamefont {Chen}}, \bibinfo
  {author} {\bibfnamefont {W.-H.}\ \bibnamefont {Wang}}, \bibinfo {author}
  {\bibfnamefont {W.}~\bibnamefont {Kob}},\ and\ \bibinfo {author}
  {\bibfnamefont {L.}~\bibnamefont {Xu}},\ }\bibfield  {title} {\bibinfo
  {title} {Topology of vibrational modes predicts plastic events in glasses},\
  }\href@noop {} {\bibfield  {journal} {\bibinfo  {journal} {Nat. Commun.}\
  }\textbf {\bibinfo {volume} {14}},\ \bibinfo {pages} {2955} (\bibinfo {year}
  {2023})}\BibitemShut {NoStop}%
\bibitem [{\citenamefont {Wojciechowski}\ and\ \citenamefont
  {Klos}(1996)}]{wojciechowski1996minimum}%
  \BibitemOpen
  \bibfield  {author} {\bibinfo {author} {\bibfnamefont {K.~W.}\ \bibnamefont
  {Wojciechowski}}\ and\ \bibinfo {author} {\bibfnamefont {J.}~\bibnamefont
  {Klos}},\ }\bibfield  {title} {\bibinfo {title} {On the minimum energy
  structure of soft, two-dimensional matter in a strong uniform field:gravity's
  rainbow'revisited},\ }\href@noop {} {\bibfield  {journal} {\bibinfo
  {journal} {J. Phys. A: Math. Gen.}\ }\textbf {\bibinfo {volume} {29}},\
  \bibinfo {pages} {3963} (\bibinfo {year} {1996})}\BibitemShut {NoStop}%
\bibitem [{\citenamefont {Mughal}\ and\ \citenamefont
  {Moore}(2007)}]{Mughal2007}%
  \BibitemOpen
  \bibfield  {author} {\bibinfo {author} {\bibfnamefont {A.}~\bibnamefont
  {Mughal}}\ and\ \bibinfo {author} {\bibfnamefont {M.}~\bibnamefont {Moore}},\
  }\bibfield  {title} {\bibinfo {title} {Topological defects in the crystalline
  state of one-component plasmas of nonuniform density},\ }\href@noop {}
  {\bibfield  {journal} {\bibinfo  {journal} {Phys. Rev. E}\ }\textbf {\bibinfo
  {volume} {76}},\ \bibinfo {pages} {011606} (\bibinfo {year}
  {2007})}\BibitemShut {NoStop}%
\bibitem [{\citenamefont {Mughal}\ and\ \citenamefont
  {Weaire}(2009)}]{Mughal2009}%
  \BibitemOpen
  \bibfield  {author} {\bibinfo {author} {\bibfnamefont {A.}~\bibnamefont
  {Mughal}}\ and\ \bibinfo {author} {\bibfnamefont {D.}~\bibnamefont
  {Weaire}},\ }\bibfield  {title} {\bibinfo {title} {Curvature in conformal
  mappings of two-dimensional lattices and foam structure},\ }\href@noop {}
  {\bibfield  {journal} {\bibinfo  {journal} {Proc. R. Soc. London, Ser. A}\
  }\textbf {\bibinfo {volume} {465}},\ \bibinfo {pages} {219} (\bibinfo {year}
  {2009})}\BibitemShut {NoStop}%
\bibitem [{\citenamefont {Yao}\ and\ \citenamefont {Olvera de~la
  Cruz}(2013)}]{Yao2013a}%
  \BibitemOpen
  \bibfield  {author} {\bibinfo {author} {\bibfnamefont {Z.}~\bibnamefont
  {Yao}}\ and\ \bibinfo {author} {\bibfnamefont {M.}~\bibnamefont {Olvera de~la
  Cruz}},\ }\bibfield  {title} {\bibinfo {title} {Topological defects in flat
  geometry: the role of density inhomogeneity},\ }\href@noop {} {\bibfield
  {journal} {\bibinfo  {journal} {Phys. Rev. Lett.}\ }\textbf {\bibinfo
  {volume} {111}},\ \bibinfo {pages} {115503} (\bibinfo {year}
  {2013})}\BibitemShut {NoStop}%
\bibitem [{\citenamefont {Soni}\ \emph {et~al.}(2018)\citenamefont {Soni},
  \citenamefont {G{\'o}mez},\ and\ \citenamefont {Irvine}}]{Soni2018}%
  \BibitemOpen
  \bibfield  {author} {\bibinfo {author} {\bibfnamefont {V.}~\bibnamefont
  {Soni}}, \bibinfo {author} {\bibfnamefont {L.~R.}\ \bibnamefont
  {G{\'o}mez}},\ and\ \bibinfo {author} {\bibfnamefont {W.~T.}\ \bibnamefont
  {Irvine}},\ }\bibfield  {title} {\bibinfo {title} {Emergent geometry of
  inhomogeneous planar crystals},\ }\href@noop {} {\bibfield  {journal}
  {\bibinfo  {journal} {Phys. Rev. X}\ }\textbf {\bibinfo {volume} {8}},\
  \bibinfo {pages} {011039} (\bibinfo {year} {2018})}\BibitemShut {NoStop}%
\bibitem [{\citenamefont {Silva}\ \emph {et~al.}(2020)\citenamefont {Silva},
  \citenamefont {Menezes}, \citenamefont {Cabral},\ and\ \citenamefont
  {de~Souza~Silva}}]{silva2020formation}%
  \BibitemOpen
  \bibfield  {author} {\bibinfo {author} {\bibfnamefont {F.~C.}\ \bibnamefont
  {Silva}}, \bibinfo {author} {\bibfnamefont {R.~M.}\ \bibnamefont {Menezes}},
  \bibinfo {author} {\bibfnamefont {L.~R.}\ \bibnamefont {Cabral}},\ and\
  \bibinfo {author} {\bibfnamefont {C.~C.}\ \bibnamefont {de~Souza~Silva}},\
  }\bibfield  {title} {\bibinfo {title} {Formation and stability of conformal
  spirals in confined 2d crystals},\ }\href@noop {} {\bibfield  {journal}
  {\bibinfo  {journal} {J. Phys.: Condens. Matter}\ }\textbf {\bibinfo {volume}
  {32}},\ \bibinfo {pages} {505401} (\bibinfo {year} {2020})}\BibitemShut
  {NoStop}%
\bibitem [{\citenamefont {Yao}(2024)}]{yao2024intrinsic}%
  \BibitemOpen
  \bibfield  {author} {\bibinfo {author} {\bibfnamefont {Z.}~\bibnamefont
  {Yao}},\ }\bibfield  {title} {\bibinfo {title} {Intrinsic conformal order
  revealed in geometrically confined long-range repulsive particles},\
  }\href@noop {} {\bibfield  {journal} {\bibinfo  {journal} {Europhys. Lett.}\
  }\textbf {\bibinfo {volume} {146}},\ \bibinfo {pages} {46003} (\bibinfo
  {year} {2024})}\BibitemShut {NoStop}%
\bibitem [{\citenamefont {Hohenberg}(1967)}]{hohenberg1967existence}%
  \BibitemOpen
  \bibfield  {author} {\bibinfo {author} {\bibfnamefont {P.~C.}\ \bibnamefont
  {Hohenberg}},\ }\bibfield  {title} {\bibinfo {title} {Existence of long-range
  order in one and two dimensions},\ }\href@noop {} {\bibfield  {journal}
  {\bibinfo  {journal} {Phys. Rev.}\ }\textbf {\bibinfo {volume} {158}},\
  \bibinfo {pages} {383} (\bibinfo {year} {1967})}\BibitemShut {NoStop}%
\bibitem [{\citenamefont {Mermin}\ and\ \citenamefont {Wagner}(1966)}]{MW1966}%
  \BibitemOpen
  \bibfield  {author} {\bibinfo {author} {\bibfnamefont {N.~D.}\ \bibnamefont
  {Mermin}}\ and\ \bibinfo {author} {\bibfnamefont {H.}~\bibnamefont
  {Wagner}},\ }\bibfield  {title} {\bibinfo {title} {Absence of ferromagnetism
  or antiferromagnetism in one- or two-dimensional isotropic heisenberg
  models},\ }\href@noop {} {\bibfield  {journal} {\bibinfo  {journal} {Phys.
  Rev. Lett.}\ }\textbf {\bibinfo {volume} {17}},\ \bibinfo {pages} {1133}
  (\bibinfo {year} {1966})}\BibitemShut {NoStop}%
\bibitem [{\citenamefont {Coleman}(1973)}]{coleman1973there}%
  \BibitemOpen
  \bibfield  {author} {\bibinfo {author} {\bibfnamefont {S.}~\bibnamefont
  {Coleman}},\ }\bibfield  {title} {\bibinfo {title} {There are no goldstone
  bosons in two dimensions},\ }\href@noop {} {\bibfield  {journal} {\bibinfo
  {journal} {Commun. Math. Phys.}\ }\textbf {\bibinfo {volume} {31}},\ \bibinfo
  {pages} {259} (\bibinfo {year} {1973})}\BibitemShut {NoStop}%
\end{thebibliography}

%

\end{document}